\documentclass{emulateapj}

\newcommand{\unit}[1]{\nobreak{\mathrm{\;#1}}} 

\newcommand{\bmath}[1]{\mbox{\boldmath{$#1$}}}

\def\comp{\,c/\omega_{\rm p}}
\def\ompt{\omega_{\rm p}t}

\newcommand{\gb}[1]{\gamma\beta_{\rm {#1}}}

\newcommand{\eq}[1]{eq.~(\ref{eq:#1})}
\newcommand{\fig}[1]{Fig.~\ref{fig:#1}}
\newcommand{\fign}[1]{\ref{fig:#1}}
\newcommand{\tit}[1]{\textit{#1}}
\newcommand{\be}{\begin{eqnarray}}
\newcommand{\ee}{\end{eqnarray}}

\begin{document}
\title{Acceleration of Particles at the Termination Shock of a Relativistic Striped Wind}
\author{Lorenzo Sironi and Anatoly Spitkovsky}
\affil{Department of Astrophysical Sciences, Princeton University, Princeton, NJ 08544-1001, USA}
\email{lsironi@astro.princeton.edu;\\ anatoly@astro.princeton.edu}
\begin{abstract}
The relativistic wind of obliquely-rotating pulsars consists of toroidal stripes of opposite magnetic field polarity, separated by current sheets of hot plasma. By means of two- and three-dimensional particle-in-cell simulations, we investigate particle acceleration and magnetic field dissipation at the termination shock of a relativistic striped wind. At the shock, the flow compresses and the alternating fields annihilate by driven magnetic reconnection. 
Irrespective of the stripe wavelength $\lambda$ or the wind magnetization $\sigma$ (in the regime $\sigma\gg1$ of magnetically-dominated flows), shock-driven reconnection transfers all the magnetic energy of alternating fields to the particles, whose average Lorentz factor increases by a factor of $\sigma$ with respect to the pre-shock value. The shape of the post-shock spectrum depends primarily on the ratio $\lambda/(r_L\sigma)$, where $r_L$ is the relativistic Larmor radius in the wind. The spectrum becomes broader as the value of $\lambda/(r_L\sigma)$ increases, passing from a relativistic Maxwellian to a flat power-law tail with slope around $-1.5$, populated by particles accelerated by the reconnection electric field. Close to the equatorial plane of the wind, where the stripes are symmetric, the highest energy particles resulting from magnetic reconnection can escape ahead of the shock, and be injected into a Fermi-like acceleration process. In the post-shock spectrum, they populate a power-law tail with slope around $-2.5$, that extends beyond the flat component produced by reconnection. Our study suggests that the spectral break between the radio and the optical band in Pulsar Wind Nebulae can be a natural consequence of particle acceleration at the termination shock of striped pulsar winds.
\end{abstract}

\section{Introduction}\label{sec:intro}
The broadband spectrum of Pulsar Wind Nebulae (PWNe), extending from the radio up to the gamma-ray band, is usually modeled as synchrotron and inverse Compton radiation from a nonthermal population of electron-positron pairs. It is assumed that the emitting particles are accelerated to a power-law distribution at the so-called termination shock, where the momentum flux of the relativistic pulsar wind is balanced by the confining pressure of the nebula.  However, it is still unclear how the wind termination shock can accelerate electrons and positrons (hereafter, simply ``electrons'') to the relativistic energies required to power the nebular emission.

A flat radio spectrum is the common observational feature of PWNe, whose spectral flux in the radio band scales as $F_{\nu}\propto\nu^{-\alpha_r}$, with $\alpha_r=0.0-0.3$. 
For the Crab Nebula, the best studied example of a PWN, the synchrotron cooling time of radio-emitting electrons largely exceeds the age of the nebula, so one cannot exclude that such particles may be ``primordial,'' i.e., accelerated at a very early stage of the PWN evolution  \citep{atoyan_99}. However, observations of wisps in the radio band \citep[][]{bietenholz_01} seem to suggest that radio-emitting electrons are being accelerated now, in the same region as those responsible for the optical and X-ray emission. In this case, the observed radio spectral index \citep[$\alpha_r\sim0.25$ for the Crab;][]{bietenholz_97} would imply a distribution of emitting electrons of the form $dN/dE\propto E^{-p}$, with slope $p=2\alpha_r+1\sim1.5$. To explain the radio through optical emission of the Crab, the power law of shock-accelerated electrons should span at least three decades in particle energy \citep[e.g.,][]{lyubarsky_03}. Flat electron spectra with slope $p<2$ below GeV energies are also required to model the radio emission of hotspots in radio galaxies \citep[][]{stawarz_07} and X-ray observations of luminous blazar sources \citep[][]{sikora_09}.

For a power-law particle spectrum with $1<p<2$, most of the particles lie at the low-energy end of the distribution, but the plasma energy content is dominated by particles at the high-energy cutoff. This  contrasts with the energy spectrum expected from Fermi acceleration in relativistic shocks, which normally yields slopes $p>2$ \citep[e.g.,][]{achterberg_01, keshet_waxman_05}, in which case low-energy particles dominate both by number and by energy. To produce broad spectra with $p<2$, a mechanism is required that raises the mean electron energy, but leaves most of the particles at relatively low energies. In the context of pulsar winds, \citet{hoshino_92} and \citet{amato_arons_06} have proposed that, in a wind loaded with ions, resonant absorption of ion cyclotron waves by electrons and positrons could generate flat distributions downstream from the termination shock. However, the required ion injection rate largely exceeds what is expected from standard models of pulsar magnetospheres \citep[e.g.,][]{arons_07}. 

An alternative possibility was discussed by \citet{lyubarsky_03}, under the assumption that the flow upstream of the termination shock consists of alternating stripes of opposite magnetic polarity, separated by current sheets of hot plasma (from now on, a ``striped wind''). For obliquely-rotating pulsars, this is the configuration expected around the equatorial plane of the wind, where the sign of the toroidal field alternates with the pulsar period.  If the wind remains dominated by Poynting flux till the termination shock \citep[which is still an open question, see][]{lyubarsky_kirk_01, kirk_sk_03}, shock-compression of the stripes may drive annihilation of the alternating fields, a process  known as ``driven magnetic reconnection.'' The resulting transfer of energy from the fields to the particles could potentially generate flat electron distributions, with slope $1<p<2$. 

The physics of magnetic reconnection can be captured self-consistently only by means of multi-dimensional particle-in-cell (PIC) simulations. Fully-kinetic PIC simulations provide a powerful tool to explore the microphysics of collisionless plasmas, since they can capture from first principles the fundamental interplay between charged particles and electromagnetic fields. In the context of relativistic magnetic reconnection in pair plasmas, most studies have explored the so-called ``undriven reconnection'' process \citep{zenitani_01,zenitani_05, zenitani_07,zenitani_08, jaroschek_04}, where field annihilation is initiated by a transient seed perturbation to an otherwise stable current sheet. As discussed above, this is not the setup expected at the wind termination shock. Here, it is the shock-compression of the flow that steadily drives regions of opposite magnetic field polarity toward each other, causing reconnection. 

Shock-driven reconnection in magnetically-dominated flows has been studied by \citet{petri_lyubarsky_07} with analytical methods and one-dimensional (1D) PIC simulations.\footnote{A similar 1D study, but for winds dominated by kinetic (rather than Poynting) flux, was performed by \citet{nagata_08}.} They find a condition on the wind magnetization and the stripe wavelength needed for full dissipation of the alternating fields at the termination shock (roughly speaking, the post-shock Larmor radius should be larger than the separation between current sheets). However, in 1D all particles gain energy at the same rate, which results in a Maxwellian distribution downstream from the shock. The generation of nonthermal tails is possible only if different particles experience different energy gains, which requires multi-dimensional simulations. As a preliminary step toward multi-dimensional studies of the termination shock in a striped wind, \citet{lyubarsky_liverts_08} have performed 2D simulations of driven magnetic reconnection. Two stripes of opposite magnetic polarity were compressed  by means of an external force, which would imitate the effect of a shock. They found that driven magnetic reconnection can produce flat nonthermal tails, with $p\sim1$. Yet, their study was limited to a single current sheet, and most importantly it did not self-consistently account for the presence of a shock.

In this work, we explore via multi-dimensional PIC simulations the acceleration of particles at the termination shock of a striped relativistic electron-positron wind. The physics of a shock interacting with multiple current sheets is captured self-consistently in 2D and 3D, and the transfer of energy from the field to the particles via shock-driven reconnection is investigated from first principles. We find that the alternating fields are completely dissipated upon compression by the shock, and their energy is transferred to the particles, regardless of the properties of the flow. This contrasts with the conclusions of \citet{petri_lyubarsky_07}, who found efficient field dissipation only for high magnetizations and short stripe wavelengths. We show that the reduced dimensionality of their 1D model could not correctly capture the growth of the so-called ``tearing-mode instability,'' which is of paramount importance for field annihilation in 2D and 3D configurations. We find that broad particle spectra with flat slopes ($1<p<2$) are a common by-product of shock-driven reconnection, but the extent of the power-law tail depends on the wind magnetization and the stripe wavelength.

This work is organized as follows. In \S \ref{sec:setup}, we discuss the setup of our simulations and the magnetic field geometry. In \S \ref{sec:struct}, the properties of the termination shock are investigated for one representative choice of wind parameters. We discuss both the shock structure, with focus on the physics of shock-driven magnetic reconnection, and the process of particle acceleration to nonthermal energies. In \S \ref{sec:cond}, we study how the particle spectrum downstream from the shock depends on the physical conditions in the wind, like magnetization and stripe wavelength. Although we mostly employ 2D simulations, in \S\ref{app:3d} we compare 2D and 3D results, finding good agreement. We summarize our findings in \S \ref{sec:disc} and comment on the application of our results to astrophysical scenarios.

\section{Simulation Setup}\label{sec:setup}
\begin{figure}[tbp]
\begin{center}
\includegraphics[width=0.5\textwidth]{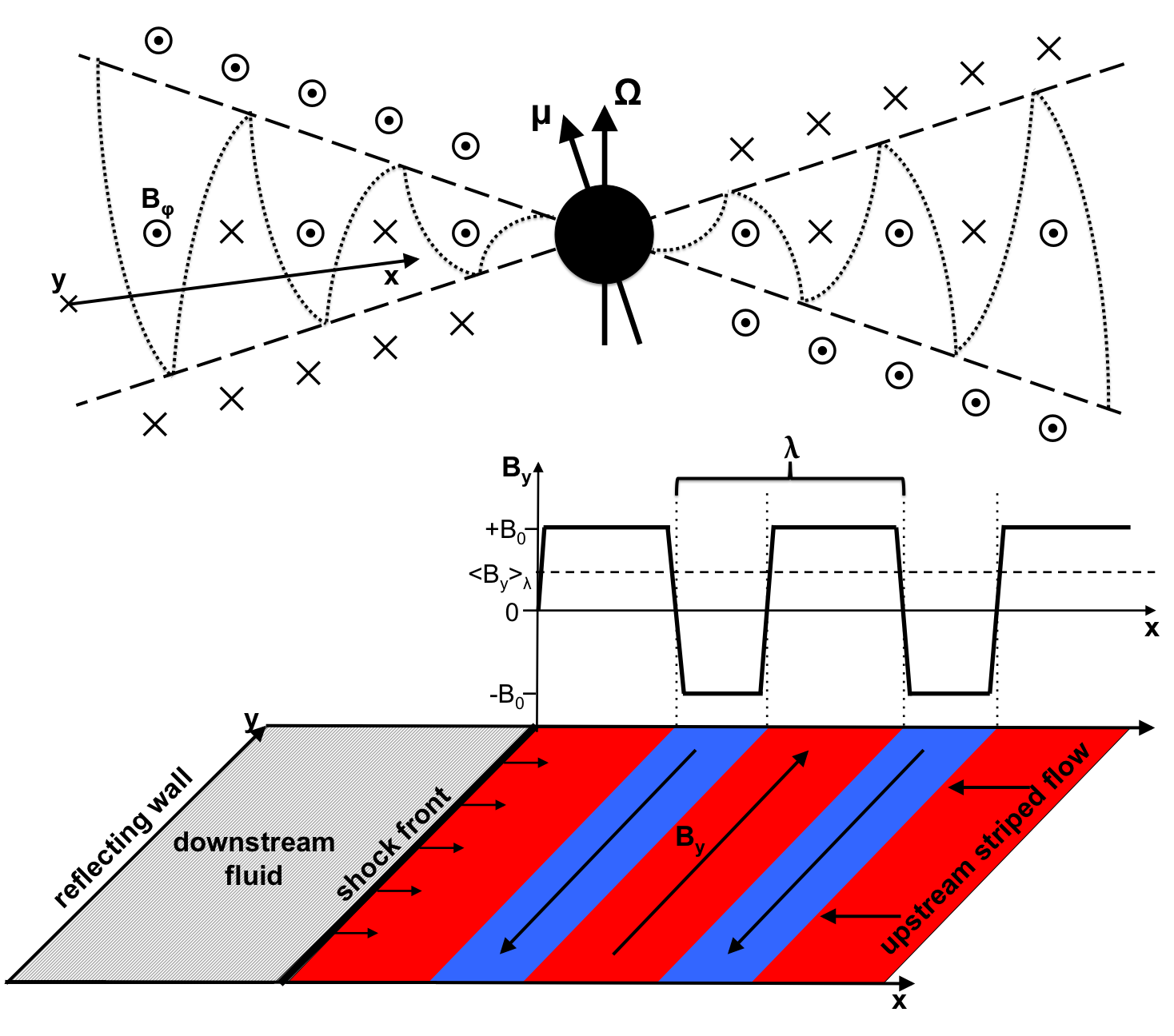}
\caption{Upper panel: poloidal structure of the striped pulsar wind, according to the solution by \citet{bogovalov_99}. The arrows denote the pulsar rotational axis (along $\bmath{\Omega}$, vertical) and magnetic axis (along $\bmath{\mu}$, inclined). Within the equatorial wedge bounded by the dashed lines, the wind consists of toroidal stripes of alternating polarity (see the reversals of $B_\varphi$), separated by current sheets (dotted lines). At latitudes higher than the inclination angle between $\bmath{\Omega}$ and $\bmath{\mu}$ (i.e., beyond the dashed lines), the field does not alternate. Lower panel: simulation geometry. For 2D runs, the simulation domain is in the $xy$ plane, oriented as shown in the upper panel (so, $\bmath{\hat{x}}=-\bmath{\hat{r}}$ and $\bmath{\hat{y}}=-\bmath{\hat{\varphi}}$). The incoming flow propagates along $-\bmath{\hat{x}}$, and the shock moves away from the reflecting wall (located at $x=0$) toward $+\bmath{\hat{x}}$. The magnetic field lies in the simulation plane along the $y$ direction, and its polarity alternates with wavelength $\lambda$ (red stripe for $B_y=+B_0$, blue for $B_y=-B_0$). A net stripe-averaged field $\langle B_y\rangle_\lambda>0$ is set up by choosing red stripes wider than the blue stripes. For the pulsar wind sketched above, $\langle B_y\rangle_\lambda>0$ is realized below the equator. }
\label{fig:simplane}
\end{center}
\end{figure}

We use the 3D electromagnetic PIC code TRISTAN-MP \citep{buneman_93, spitkovsky_05} to study the termination shock of a relativistic striped wind. The shock is set up by reflecting a magnetized electron-positron  flow from a conducting wall located at $x = 0$ (\fig{simplane}). The interaction between the incoming beam (that propagates along $-\bmath{\hat{x}}$) and the reflected beam triggers the formation of a shock, which moves away from the wall along $+\bmath{\hat{x}}$ (\fig{simplane}). The simulation is performed in the ``wall'' frame, where the downstream plasma is at rest.

To follow the shock evolution for longer times with fixed computational resources, we mainly utilize 2D computational domains in the $xy$ plane, with periodic boundary conditions in the $y$ direction. In \S\ref{app:3d} we compare 2D and 3D simulations and show that 2D runs can capture most of the relevant physics. For both 2D and 3D domains, all three components of particle velocities and electromagnetic fields are tracked.

The incoming electron-positron stream is injected along $-\bmath{\hat{x}}$ with bulk Lorentz factor $\gamma_0$. The flow carries a strong magnetic field, oriented along the $y$ direction. It follows that the shock will be ``perpendicular,'' i.e., with magnetic field orthogonal to the shock normal. The choice to initialize the field in the simulation plane (i.e., oriented along $y$, as opposed to $z$) is motivated by the agreement between 2D simulations with in-plane fields and 3D experiments, as we show in \S\ref{app:3d}. The spatial profile of the injected magnetic field, sketched in \fig{simplane}, is 
\be\label{eq:by}
\!\!\!B_{y}(x,t)&=&B_0 \tanh\left\{\frac{1}{\delta}\left[\alpha+\cos\left(\frac{2\pi(x+\beta_0 c t)}{\lambda}\right)\right]\right\}\\\nonumber
&\equiv& B_0 \tanh\zeta
\ee
and its associated motional electric field is $E_z=\beta_0 B_y$. Here, $\beta_0$ is the three-velocity of the injected plasma. This particular choice for the magnetic field is motivated by analytical studies of the structure of pulsar winds \citep[e.g.,][]{petri_kirk_05}. At each given time, \eq{by} shows that the field profile is periodic with wavelength $\lambda$ (see \fig{simplane}). Current sheets, where the magnetic field vanishes, are identified by the condition $\zeta\simeq0$, and their half-thickness is $\Delta=\lambda\delta/2\pi\ll\lambda$. Across each current sheet, the field reverses its sign from $+B_0$ to $-B_0$ (or vice versa), where $B_0$ is the field strength in the region outside the current sheets, which we shall call ``cold wind.'' In the following, we parameterize the field strength via the so-called magnetization parameter $\sigma\equiv B_{0}^2/4 \pi \gamma_0 m n_{c0} c^2$, taken to be much larger than unity. Here, $m$ is the electron (or positron) mass and $n_{c0}$ is the density of particles  in the cold wind. By defining the relativistic skin depth in the cold wind $c/\omega_{\rm p}\equiv \sqrt{\gamma_0 m c^2/ 4\pi e^2 n_{c0}}$, and the relativistic Larmor radius $r_{L}\equiv \gamma_0 mc^2/eB_0$, we can rewrite the magnetization parameter as $\sigma=(c/\omega_{\rm p})^2/r_{L}^2$. Finally, the parameter $\alpha$ ($-1<\alpha<1$) is a measure of the net field (i.e., averaged over one wavelength $\lambda$), such that $\langle B_y\rangle_\lambda/B_0=\alpha/(2-|\alpha|)$. Although the magnetic field intensity in the cold wind is always $B_0$, the wavelength-averaged field is not necessarily zero, because two neighboring stripes generally have different widths (see \fig{simplane}). In the pulsar wind, one expects $\langle B_y\rangle_\lambda=0$ (which corresponds to $\alpha=0$) only in the equatorial plane (see upper panel in \fig{simplane}). We remark that a net field $\langle B_y\rangle_\lambda\neq0$ does not play the role of a guide field, which in our geometry would correspond to a uniform component of magnetic field in the $z$ direction. No significant guide field is expected at the termination shock of pulsar winds, and the physics of shock-driven reconnection in the presence of a guide field will be discussed elsewhere.

In the cold wind, each computational cell is initialized with two electrons and two positrons, with a small thermal spread $\Theta_c\equiv k T_c/m c^2=10^{-4}$. We have also performed limited experiments with a larger number of particles per cell (up to 32 per species), obtaining essentially the same results. Such a background of cold particles is initialized also within the current sheets, but with the addition of a hot component whose density profile is $n_h=n_{h0}/\cosh^2\zeta$. Here $n_{h0}\equiv\eta\, n_{c0}$ is the density of hot particles at the center of the current sheet. The temperature of this hot component is determined by pressure balance between the gaseous part (inside the current sheet) and the magnetic part (in the cold wind), which yields a thermal spread $\Theta_h\equiv k T_h/mc^2=\sigma/2\eta$ for a 3D relativistic Maxwellian. Finally, the variation of magnetic field across a current sheet needs to be sustained by a current flowing along  $\bmath{\hat{z}}$. In the wind frame, electrons and positrons in the current sheet should be drifting in opposite directions with the same speed $\beta_{h}=\sqrt{\sigma}/(\eta\gamma_0)\,(c/\omega_{\rm p})/\Delta$. In summary, to initialize the distribution of hot electrons (and positrons) in the frame of the simulations, we need to make two successive Lorentz boosts: first, from their proper frame to the wind frame, where electrons and positrons counter-stream along $\bmath{\hat{z}}$ with velocity $\pm\beta_h$; then, from the wind frame to the simulation frame, where the wind propagates along $-\bmath{\hat{x}}$ with Lorentz factor $\gamma_0$.\footnote{In the range of parameters we explore, we find that $\beta_h\ll1$, so the first Lorentz boost can be safely neglected.}

In our study, we adopt $\gamma_0=15$ as our fiducial bulk Lorentz factor. As discussed in \S \ref{sec:cond}, we actually explore a wide range of Lorentz factors, from $\gamma_0=3$ to $\gamma_0=375$, and find basically the same results, modulo an overall shift in the energy scale. We vary the stripe wavelength by almost two orders of magnitude, from $\lambda=20\comp$ to $\lambda=1280\comp$, and we discuss its effects on the structure of the shock and the process of shock-driven reconnection in \S \ref{sec:lambda}. The dependence on the wind magnetization, though always in the regime of highly magnetized flows ($\sigma\gg1$), is presented in \S \ref{sec:sigma}, where we vary $\sigma$ from 10 to 100. Different values of the stripe-averaged magnetic field are investigated in \S \ref{sec:alpha}, where we vary the parameter $\alpha=2\langle B_y\rangle_\lambda/(B_0+|\langle B_y\rangle_\lambda|)$ between 0.0 and 0.95. For pulsar winds, large values of $|\alpha|$ would be expected at high latitudes above the midplane, whereas $\alpha=0$ along the equator (see upper panel in \fig{simplane}). 

The structure of the current sheet is completely determined once we fix the value of the density excess $\eta=n_{h0}/n_{c0}$, and of the sheet half-thickness $\Delta=\lambda\delta/2\pi$. We choose $\eta=3$,  motivated by analytical studies of pulsar winds \citep{lyubarsky_kirk_01}. However, we find that our results do not appreciably depend on the value of $\eta$ (we have also tried with $\eta=1.5$ and $\eta=6$), provided that the contribution of current sheets to the upstream particle flux stays negligible (i.e., $\eta\delta\ll1$). Regarding the sheet half-thickness, we typically employ $\Delta=\comp$ (for each choice of $\lambda$, this determines the value of $\delta=2\pi \Delta/\lambda$), but we have verified that our results are not very sensitive to this parameter (we have tried up to $\Delta=8\comp$, for $\lambda=320\comp$). The choice for the sheet thickness is guided, on the one hand, by the requirement that the contribution of current sheets to the upstream plasma energy flux should be small (i.e., $\sigma\delta\ll1$). On the other hand, we enforce  $\Delta\gtrsim0.5\comp$, to ensure that the incoming current sheets are stable to the process of \textit{undriven} reconnection. With this choice, we are confident that the trigger for magnetic dissipation will come entirely from the \textit{shock-driven} compression of the flow.

We choose the relativistic electron (or positron) skin depth in the cold wind $c/\omega_{\rm p}$ such that the smallest scale in the system, which for $\sigma>1$ is the relativistic Larmor radius $r_{L}=(c/\omega_{\rm p})/\sqrt{\sigma}$, is resolved with at least a few computational cells.\footnote{The skin depth and Larmor radius for the hot particles in the current sheet are larger than their counterparts in the cold wind by a factor of $\sqrt{\sigma/2\eta^2}$ and $\sigma/2\eta$, respectively.} We resolve $c/\omega_{\rm p}$ with 7.5 computational cells for $\sigma=10$, 10 cells for $\sigma=20$, 13 cells for $\sigma=30$, 17 cells for $\sigma=50$, and 24 cells for $\sigma=100$. For each value of the magnetization, we have verified that our results do not substantially change when doubling the resolution (for $\sigma=10$, we have tried with up to $c/\omega_{\rm p}=30$ cells). Our computational domain is typically $\sim400\comp$ wide (along the $y$ direction). In particular, we choose a transverse size of $3072$ cells for $\sigma=10$, of $4096$ cells for $\sigma=20$, of $5120$ cells for $\sigma=30$, of $7168$ cells for $\sigma=50$, and of $9216$ cells for $\sigma=100$. A large box is of paramount importance for the consistency of our findings. As we show in Appendix \ref{app:my}, the results of our simulations converge only when the computational box is larger than the half-wavelength of the stripes. It follows that 1D simulations, as the ones presented by \citet{petri_lyubarsky_07}, cannot capture all the relevant physics, especially for long wavelengths. 

The results presented in the following sections refer to shocks that have already reached a steady state. For the typical parameters we explore, this happens before $\ompt=3000$, but in some cases (especially for long stripe wavelengths), we need to evolve our simulations up to longer times. In Appendix \ref{app:time}, we follow the entire time evolution of a representative shock, from the earliest stages until it approaches a steady state.

\section{Shock Structure and Particle Acceleration}\label{sec:struct}
In this section, we present the properties of a shock that propagates in a relativistic electron-positron flow with \tit{alternating} fields (a ``striped wind''). The case of \tit{uniform} pre-shock fields has been investigated by \citet{sironi_spitkovsky_09} and \citet{sironi_spitkovsky_11a} via multi-dimensional PIC simulations (for electron-positron and electron-ion plasmas, respectively). They studied how the magnetic field strength and obliquity can affect the efficiency of the Fermi process, where particles gain energy by bouncing back and forth across the shock \citep[e.g.,][]{blandford_eichler_87}. They found that in perpendicular shocks i.e., with field orthogonal to the shock normal, the Fermi mechanism can operate only if the flow is weakly magnetized.  For magnetizations $\sigma\gtrsim10^{-3}$, the field is so strong that charged particles are forced to slide along the field lines, whose orientation prohibits repeated crossings of the shock, thus inhibiting the Fermi process.

In a medium with alternating fields, we find that the efficiency of Fermi acceleration is controlled by the magnitude of the stripe-averaged field $\langle B_y\rangle_\lambda$. Striped flows with $\langle B_y\rangle_\lambda/B_0\ll1$ are equivalent to weakly magnetized shocks in uniform fields, with respect to particle acceleration via the Fermi process. In this section, we choose a value for the stripe-averaged field $\langle B_y\rangle_\lambda/B_0\simeq0.05$ (corresponding to $\alpha=0.1$) that is large enough to inhibit the Fermi process. This allows us to focus solely on the process of shock-driven reconnection, and to isolate its effects on the shock structure and the particle spectrum. In pulsar winds, the special case $\langle B_y\rangle_\lambda=0$ should be realized only along the equatorial plane of the wind. By choosing $\alpha=0.1$, we address the generic properties of the flow above (or below) the midplane.

In this section, we adopt $\gamma_0=15$ and $\sigma=10$ as our fiducial values for the upstream bulk Lorentz factor and magnetization. To clarify the physics of shock-driven reconnection, we employ  relatively large values for the stripe wavelength ($\lambda=320\comp$ or $\lambda=640\comp$), and we choose $\alpha=0.1$, corresponding to a stripe-averaged field $\langle B_y\rangle_\lambda/B_0\simeq0.05$. The dependence of the shock properties on $\gamma_0$, $\sigma$,  $\lambda$, and $\alpha$ is investigated in \S\ref{sec:cond}. There, we also explore the special case $\alpha=0$, where shock-driven reconnection coexists with  Fermi acceleration.

\subsection{Shock Structure}\label{sec:shock}
\begin{figure}[tbp]
\begin{center}
\includegraphics[width=0.5\textwidth]{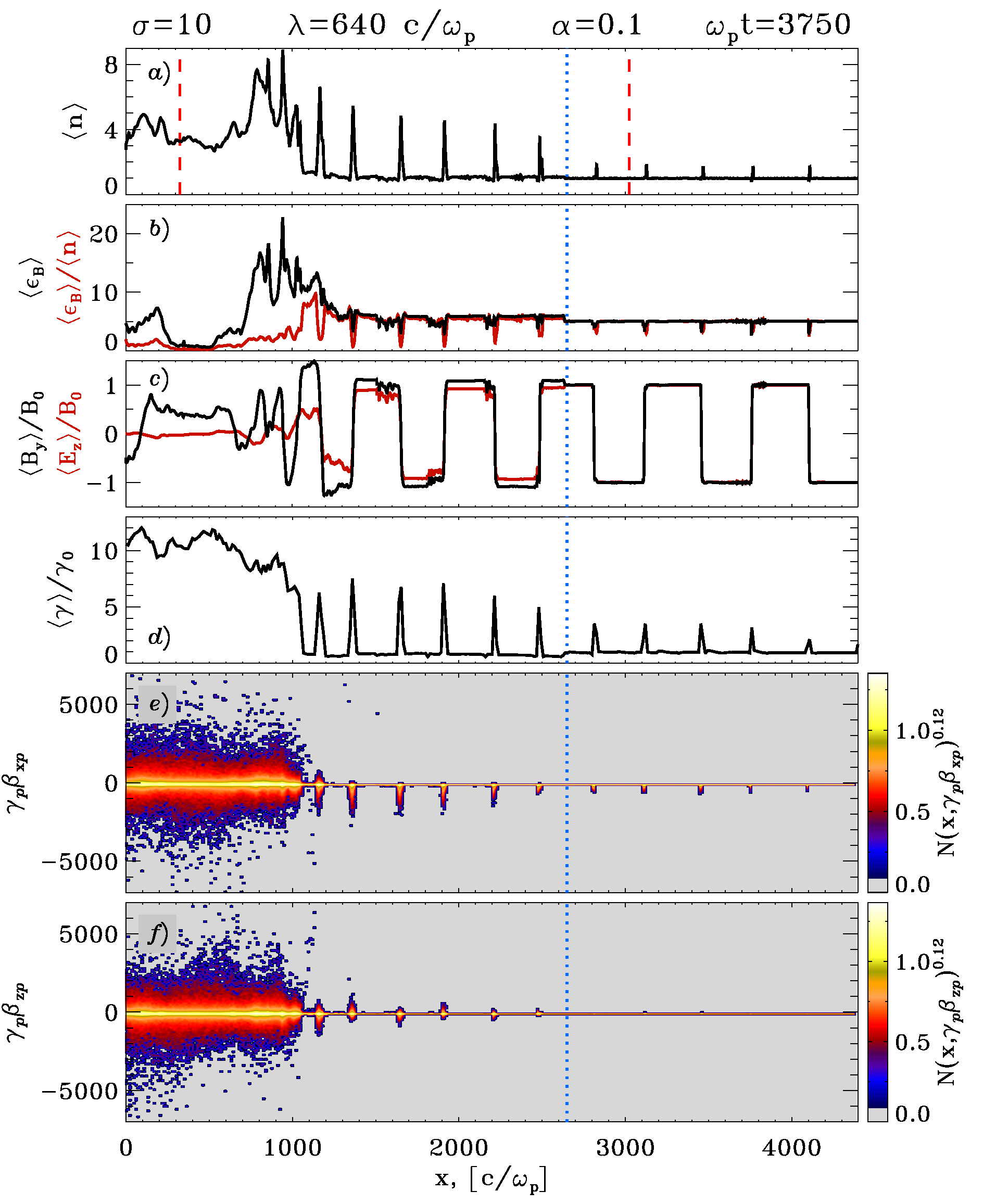}
\caption{Internal structure of the flow at $\ompt=3750$, for stripe wavelength $\lambda=640\comp$ and magnetization $\sigma=10$. The stripe-averaged field is $\langle B_y\rangle_\lambda/B_0\simeq0.05$, corresponding to $\alpha=0.1$. The hydrodynamic shock is located at $x\simeq1000\comp$. In all panels, the location of the fast MHD shock ($x\simeq2600\comp$) is indicated with a vertical dotted blue line. As a function of the longitudinal coordinate $x$, the following quantities are plotted: (a) $y$-averaged particle number density $n$, in units of the upstream value; (b) $y$-averaged magnetic energy fraction $\epsilon_B\equiv B^2/8 \pi \gamma_0 m n_{c0} c^2$ (black line), and $y$-averaged magnetic energy per particle $\epsilon_B/n$ (red line); (c) $y$-averaged magnetic field $B_y$ (black line) and electric field $E_z$ (red line), normalized to the upstream magnetic field $B_{0}$; (d) mean kinetic energy per particle, in units of the bulk energy at injection; (e) $x-\gb{x}$ positron phase space; (f) $x-\gb{z}$ positron phase space.}
\label{fig:fluidtot}
\end{center}
\end{figure}

\begin{figure*}[tbp]
\begin{center}
\includegraphics[width=\textwidth]{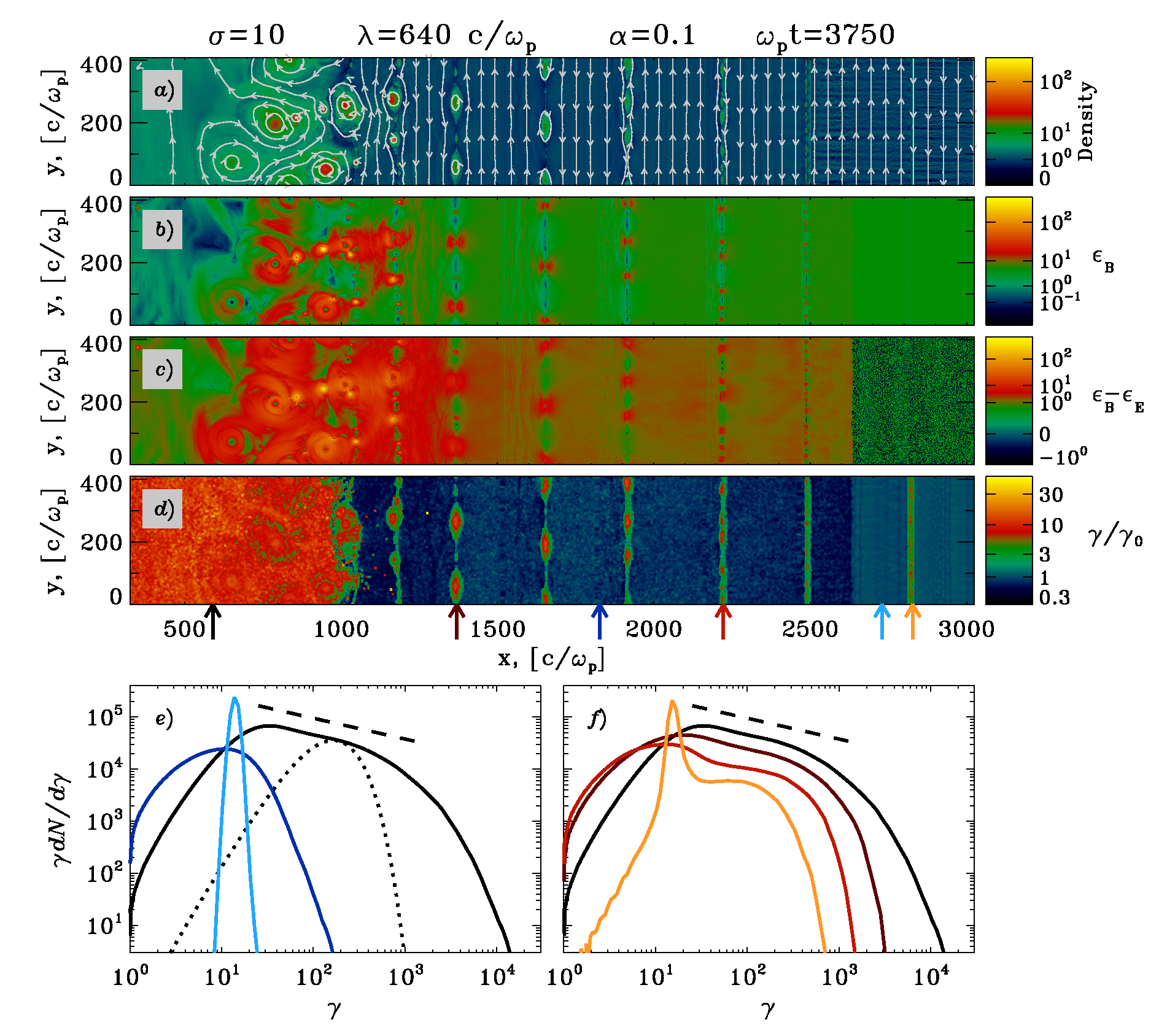}
\caption{Internal structure of the flow at $\ompt=3750$, for $\lambda=640\comp$, $\sigma=10$, and $\alpha=0.1$, zooming in on a region around the shock, as delimited by the vertical dashed red lines in \fig{fluidtot}(a). The hydrodynamic shock is located at $x\simeq1000\comp$, and the fast MHD shock at $x\simeq2600\comp$. The following quantities are plotted: (a) 2D plot in the simulation plane of the particle number density, in units of the upstream value, with contours showing the magnetic field lines; (b) 2D plot of the magnetic energy fraction $\epsilon_B\equiv B^2/8 \pi \gamma_0 m n_{c0} c^2$; (c) 2D plot of the difference $\epsilon_B-\epsilon_E$, where the electric energy fraction is $\epsilon_E\equiv E^2/8 \pi \gamma_0 m n_{c0} c^2$; (d) 2D plot of the mean kinetic energy per particle, in units of the bulk energy at injection; (e)-(f) particle energy spectra, at different locations across the flow, respectively outside (panel (e)) or inside (panel (f)) of current sheets. The color of each spectrum matches the color of the corresponding arrow at the bottom of panel (d), showing where the spectrum is computed. The dotted line in panel (e) is a Maxwellian with the same average energy as the downstream particles; the dashed lines in panels (e) and (f) indicate a power-law distribution with slope $p=1.4$.}
\label{fig:fluidsh}
\end{center}
\end{figure*}

The typical structure of a relativistic shock propagating in a striped wind is shown in Figs.~\fign{fluidtot} and \fign{fluidsh}, which refer to $\ompt=3750$, after the flow structure has reached a steady state. 
The longitudinal profile of the transition region, along the direction $x$ of shock propagation, is presented in \fig{fluidtot}, whereas \fig{fluidsh} shows the flow properties in the $xy$ plane of the simulation, zooming in on a region around the shock (as delimited by the vertical dashed red lines in \fig{fluidtot}(a)).

The incoming flow ($x\gtrsim2600\comp$) carries an alternating magnetic field of wavelength $\lambda=640\comp$ (\fig{fluidtot}(c); black line for $B_y$, red for $E_z$). The field reverses its polarity across current sheets filled with hot overdense plasma (see the spikes in density and average particle energy of \fig{fluidtot}(a) and (d), respectively), where the magnetic energy is much lower than in the cold wind (see the troughs in the magnetic energy fraction $\epsilon_B=B^2/8\pi\gamma_0mn_{c0}c^2$, black line of \fig{fluidtot}(b)). For $x\gtrsim2600\comp$, the only perturbation to the incoming flow comes from the so-called electromagnetic precursor, a coherent train of transverse electromagnetic waves propagating into the upstream at the speed of light (see the short-scale ripples at $x\simeq3850\comp$ in \fig{fluidtot}(b) and (c)). The electromagnetic precursor wave is generated in the initial stages of shock evolution, before shock-driven reconnection appreciably changes the structure of the shock. At such early times, the flow resembles a magnetized shock propagating in uniform fields (see Appendix \ref{app:time}), which is well known to radiate electromagnetic waves into the upstream  \citep[][]{gallant_92, hoshino_08, sironi_spitkovsky_09, sironi_spitkovsky_11a}.

The steady-state structure of the flow in \fig{fluidtot} shows the presence of two shocks. The main shock (which we shall call the ``hydrodynamic shock'' from now on, for reasons that will become clear below) corresponds to the jump in density occurring at $x\simeq1000\comp$. At $x\simeq2600\comp$, well ahead of the hydrodynamic shock, the incoming flow crosses a fast magnetohydrodynamic (MHD) shock, whose location is highlighted in \fig{fluidtot} by a vertical dotted blue line. At the fast shock, the wind is decelerated and compressed, as shown by the modest increase in density (\fig{fluidtot}(a)), magnetic energy (black line in \fig{fluidtot}(b)), and transverse magnetic field $B_{y}$ (black line in  \fig{fluidtot}(c)). The deceleration of the flow appears most clearly in panel (c), as a mismatch between the motional electric field $E_z$ and the magnetic field $B_y$ behind the fast shock (red and black line, respectively). Irrespective of the initial bulk Lorentz factor of the flow, the plasma drift velocity  ($=-E_z/B_y\, \bmath{\hat{x}}$) downstream from the fast shock is only moderately relativistic, since $E_z/B_y\simeq0.8$ there (\fig{fluidtot}(c)). The flow deceleration at the fast shock is accompanied by a partial isotropization of the particle distribution, mostly in the $xz$ plane orthogonal to the field. Overall, the flow kinetic energy decreases  across the fast shock from $\gamma_0=15$ to $\langle\gamma\rangle\simeq10$ (\fig{fluidtot}(d)), to compensate for the growth in magnetic energy (\fig{fluidtot}(b)).

The fast MHD shock has a dramatic impact on the structure of the striped wind. Across the fast shock, the stripe wavelength shrinks from $\lambda=640\comp$ to $\lambda\simeq550\comp$ (see \fig{fluidtot}(c)). Most importantly, the compression induced by the passage of the fast shock through a given current sheet triggers the onset of magnetic reconnection. Small-scale islands develop inside the current sheet as a result of the tearing-mode instability. The growth and evolution of reconnection islands is clearly shown in the 2D plots of density and magnetic energy of \fig{fluidsh}(a) and (b), respectively. The current sheet at $x\simeq2500\comp$, just downstream from the fast shock (located at $x\simeq2600\comp$), appears partitioned into a number of overdense islands with strong magnetic fields (red point-like regions in \fig{fluidsh}(b), at $x\simeq2500\comp$). A magnetic X-point exists in between each pair of neighboring islands, where field lines of opposite polarity break and reconnect. As the flow recedes from the fast shock, the small-scale islands created inside each current sheet coalesce to form bigger islands. This temporal evolution corresponds, in the snapshot  taken at $\ompt=3750$ in \fig{fluidsh}, to a pattern of bigger (and so, fewer) islands for current sheets that are farther behind the fast shock.

As reconnection proceeds with distance downstream from the fast shock, more and more magnetic energy is released and transferred to particles. In fact, the spikes in average particle energy of \fig{fluidtot}(d) become stronger for current sheets farther from the MHD shock (see also the longitudinal phase space of positrons in \fig{fluidtot}(e), for $1000\comp\lesssim x\lesssim 2600\comp$). The transfer of magnetic energy to the particles continues up to the point where reconnection islands fill the entire region between neighboring current sheets. Now, the striped structure of the flow is erased, and a hydrodynamic shock forms (located at $x\simeq1000\comp$ in Figs.~\fign{fluidtot} and \fign{fluidsh}). 

Behind the hydrodynamic shock, the flow comes to rest (in the wall frame of our simulations), and the particle distribution is isotropic in three dimensions (compare longitudinal and transverse phase spaces in \fig{fluidtot}(e) and (f), respectively). The post-shock number density approaches $n_{\rm d}\simeq4\,n_{\rm u}$ (\fig{fluidtot}(a), where $n_{\rm u}$ is the density ahead of the fast shock), and the shock velocity is $\beta_{\rm sh}\simeq1/3$ (see Appendix \ref{app:time}), in agreement with the jump conditions of a relativistic \tit{unmagnetized} 3D plasma. In fact, most of the energy per particle downstream from the hydrodynamic shock is in kinetic form (as opposed to the dominant electromagnetic component of the incoming striped flow). This becomes clear when comparing the profile of the mean \tit{kinetic} energy per particle (\fig{fluidtot}(d)) with the 1D plot of the mean \tit{magnetic} energy per particle (red line in \fig{fluidtot}(b)).\footnote{We point out that in the incoming flow electric and magnetic fields equally contribute to the energy balance, whereas the red line in \fig{fluidtot}(b) only includes magnetic fields. However, downstream from the hydrodynamic shock, the flow is at rest (in the simulation frame), and electric fields vanish (see red line in \fig{fluidtot}(c)).} Behind the hydrodynamic shock, the average kinetic energy per particle increases from $\langle \gamma\rangle/\gamma_0\simeq1$ up to $\langle \gamma\rangle/\gamma_0\simeq\sigma+1\simeq11$, as expected in the case of full dissipation of magnetic fields. Correspondingly, the mean magnetic energy per particle drops to zero (red line in \fig{fluidtot}(b)). As best seen in experiments with a smaller stripe wavelength,  in the case $\alpha\neq0$ a net magnetic field survives downstream from the shock, as predicted by \citet{petri_lyubarsky_07}, and shown by the 1D profile of $B_y$ in \fig{fluidtot}(c) (black line). This residual magnetic field  results from shock-compression of the stripe-averaged upstream field (we remind that $\alpha=0.1$ here, corresponding to $\langle B_{\rm y}\rangle_\lambda/B_0\simeq0.05$).

We point out that the picture described above represents accurately the long-term behavior of the shock. In particular, the distance between the fast and the hydrodynamic shock ($\simeq1600\comp$ in Figs.~\fign{fluidtot} and \fign{fluidsh}) stays approximately constant in time (see Appendix \ref{app:time}). The hydrodynamic shock moves with $\beta_{\rm sh}\simeq1/3$, and the fast MHD shock remains far enough ahead of the main shock such that to guarantee that reconnection islands will  fill the space between neighboring current sheets, by the time the flow enters the hydrodynamic shock. For a fixed reconnection rate, this implies that the distance between the fast and the hydrodynamic shock will scale linearly with the stripe wavelength (as will the size of reconnection islands just ahead of the main shock), a prediction that we have directly verified in our simulations.

Finally, we refer to Appendix \ref{app:my} for a detailed discussion of the differences between our 2D results and the 1D simulations presented by \citet{petri_lyubarsky_07}. As apparent in \fig{fluidsh}, the process of island coalescence, which is essential for the formation of the hydrodynamic shock, can be captured correctly only with multi-dimensional simulations. Indeed, for the parameters employed in this section, the 1D model by \citet{petri_lyubarsky_07} would predict negligible field dissipation, in sharp contrast with our findings. This clearly emphasizes the importance of multi-dimensional physics for our understanding of shock-driven reconnection.

\subsection{Particle Spectrum and Acceleration}\label{sec:accel}
The particle energy spectrum at different locations through the flow, as marked by arrows at the bottom of \fig{fluidsh}(d), is shown in \fig{fluidsh}(e) and (f). Panel (e) shows the particle distribution in the cold wind, whereas panel (f) focuses on the hot plasma within current sheets. Downstream from the hydrodynamic shock, the striped structure is completely erased, and no difference persists between current sheets and cold wind. Here, the particle spectrum (black line in both panels) is in the form of a flat power-law tail (with spectral index $p\simeq1.4$, dashed line) extending from $\gamma_{\rm min}\simeq30$ to $\gamma_{\rm max}\simeq500$, where it cuts off exponentially. Since $1<p<2$, most of the particles are found at $\gamma\simeq\gamma_{\rm min}$, but most of the energy is contributed by particles at $\gamma\simeq\gamma_{\rm max}$. For comparison, a 3D Maxwellian with the same average Lorentz factor $\langle\gamma\rangle=\gamma_0(\sigma+1)\simeq165$ is plotted in panel (e) as a dotted line, to show that the actual particle spectrum is much broader than a thermal distribution.\footnote{We remark that the particle spectrum behind the hydrodynamic shock does not significantly evolve with time, after the shock has reached a steady state (see Appendix \ref{app:time}).}  By following the flow from the fast to the hydrodynamic shock, we now clarify how such a particle spectrum is generated.

Most of the incoming plasma is contained in the cold striped wind exterior to current sheets, and its spectrum is initially a cold Maxwellian drifting with bulk Lorentz factor $\gamma_0=15$ (light blue line in \fig{fluidsh}(e)). After crossing the fast shock, the flow decelerates and it partly isotropizes, which explains why the particle spectrum is now broader and it peaks at lower energies (dark blue line in \fig{fluidsh}(e), but see also the 2D plot of average particle energy in \fig{fluidsh}(d), and its 1D projection in \fig{fluidtot}(d)). 

The evolution of the particle population within current sheets is much more dramatic, as shown in \fig{fluidsh}(f).\footnote{We remind that the spatial pattern shown in \fig{fluidsh}, as a function of distance behind the MHD shock, corresponds to the time history of a given fluid element after its crossing of the fast shock.} Ahead of the fast shock, the spectrum at $x\simeq2850\comp$  (yellow line in panel (f)) shows, in addition to the cold plasma in the striped wind, the hot particles in the current sheet. Their distribution is initialized as a drifting 3D Maxwellian with thermal spread $\Theta_{h}=\sigma/2\eta$, so that in the simulation frame the spectrum should peak at $\langle \gamma \rangle=3\gamma_0\Theta_h\simeq75$, as observed (we remind that $\sigma=10$ and $\eta=3$). Behind the fast shock (red line for $x\simeq2200\comp$, brown line for $x\simeq1350\comp$), the particle spectrum consists of two components: a low-energy peak, reminiscent of the particle distribution far from current sheets (compare with the dark blue line in \fig{fluidsh}(e)); and a high-energy component, that grows in number and extends to higher and higher energies, as the flow propagates toward the hydrodynamic shock. This component is populated by particles that were initially outside the current sheet, and have entered the sheet in the course of the reconnection process. They mostly end up in the reconnection islands of \fig{fluidsh}, that act as reservoirs of particles (\fig{fluidsh}(a)) and particle energy (the high-energy regions of panel (d), colored in red, match very well the overdense islands of panel (a)). In fact, the main contributions to the high-energy end of the spectrum at $x\simeq1350\comp$ (brown line in \fig{fluidsh}(f)) come from the two major islands of that current sheet.

As magnetic reconnection proceeds with distance behind the fast shock, particles are accelerated to higher and higher energies, as we discuss below in more detail. This continues until reconnection islands grow as big as the distance between neighboring sheets, at which point the current sheets occupy the entire flow (and the hydrodynamic shock forms). This explains the smooth transition from the spectrum within current sheets to the distribution downstream of the hydrodynamic shock (yellow through black lines in \fig{fluidsh}(f)). Also, since most of the particles that end up with high energies were initially part of the cold wind, the resulting spectrum behind the hydrodynamic shock is almost independent of the initial setup of the current sheet (most notably, sheet thickness $\Delta$ and overdensity $\eta$).

\begin{figure}[tbp]
\begin{center}
\includegraphics[width=0.5\textwidth]{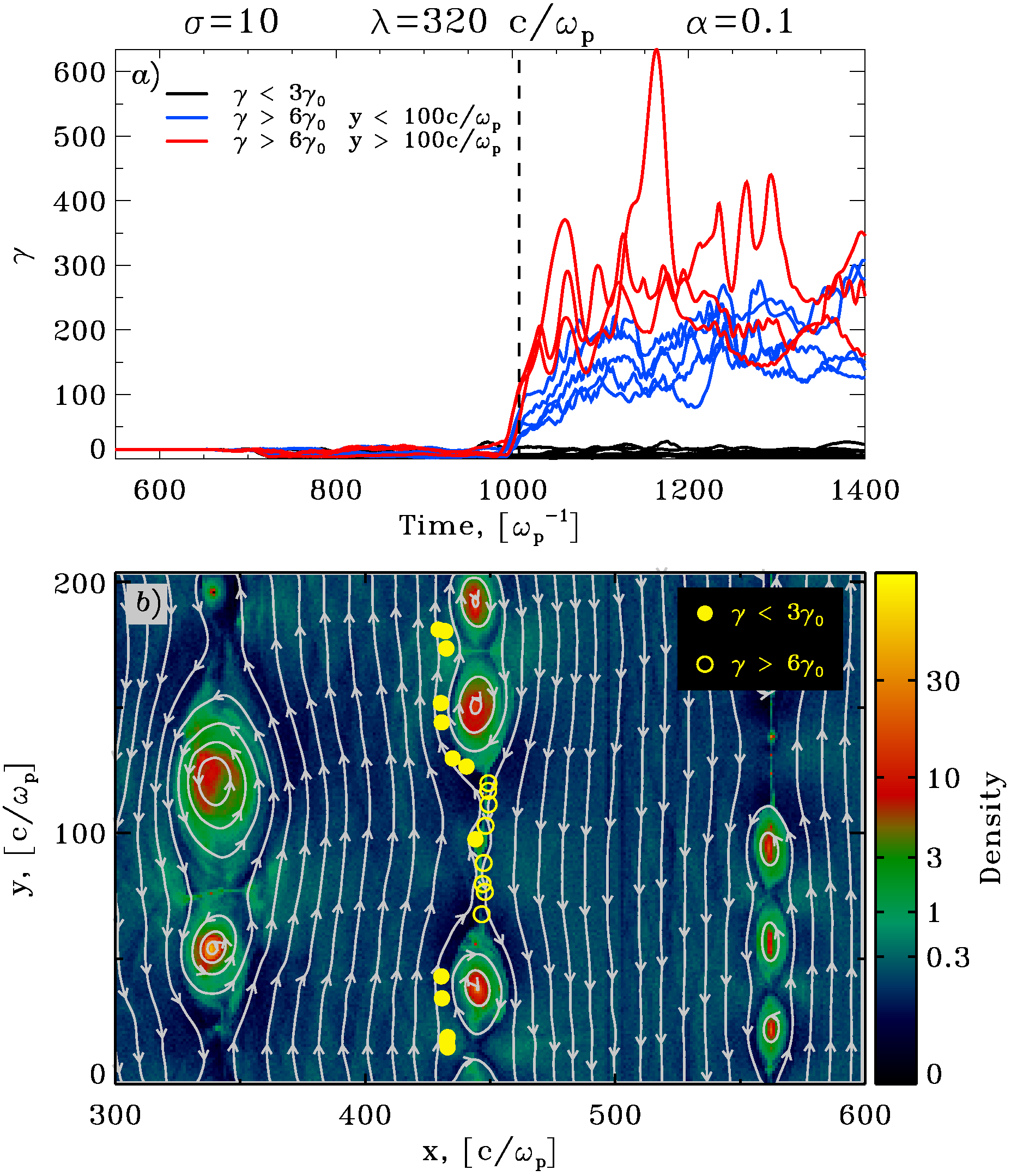}
\caption{Energy evolution of a sample of representative positrons interacting with a current sheet, in a striped flow with $\lambda=320\comp$, $\sigma=10$, and $\alpha=0.1$. All the selected positrons have been initialized as part of the cold wind, at roughly the same $x$-location. Panel (a): Energy histories of the selected positrons, with lines of different color depending on their final energy (black if $\gamma<3\gamma_0$, blue or red if $\gamma>6\gamma_0$) and their $y$-location at $\ompt=1008$ (blue if $y<100\comp$, red if $y>100\comp$). Panel (b): Magnetic field lines (white contours) superimposed on the 2D plot of particle number density, at time $\ompt=1008$ (marked as a vertical dashed line in panel (a)); the locations of the selected positrons are shown as yellow filled or open circles, depending on their final energy (filled circles for particles with $\gamma<3\gamma_0$, open circles for $\gamma>6\gamma_0$).}
\label{fig:accel}
\end{center}
\end{figure}

We now investigate in more detail the physics of particle acceleration within current sheets. In \fig{accel}, we follow the trajectories of a representative sample of positrons, extracted from the simulation of a striped wind with $\lambda=320\comp$ (but our conclusions hold for all wavelengths). All the selected positrons are initially in the cold wind (i.e., far from current sheets), at roughly the same $x$-location (within a range of $5\comp$). At first they propagate toward the fast shock, which they cross at $\ompt\simeq700$, as shown by the spreading of their energies in \fig{accel}(a). They encounter the closest current sheet (in this case, the one that was to the right of their initial position) at $\ompt=1008$, as shown in \fig{accel}(b) together with the corresponding 2D plot of particle number density. Starting from this time, their histories can diverge significantly, depending on their $y$-location at the moment of interaction with the current sheet. The particles that will end up with relatively low energies ($\gamma<3\gamma_0$, black lines in panel (a)) are found at $\ompt=1008$ exclusively around magnetic islands (their locations are indicated with filled circles in panel (b)). At later times, they will get trapped on the outskirts of the growing islands, without appreciable changes in energy. 

On the contrary, the positrons that will eventually reach high energies ($\gamma>6\gamma_0$, red and blue lines in  \fig{accel}(a)) are concentrated at $\ompt=1008$ in the vicinity of magnetic X-points (the particle locations are shown with open circles in \fig{accel}(b)). It is at X-points that magnetic field lines of opposite polarity are converging and tearing (white contours in \fig{accel}(b) show the magnetic field lines), and magnetic energy is transferred to particles. Starting from $\ompt=1008$ (vertical dashed line in panel (a)), the energies of these positrons grow simultaneously and explosively (red and blue lines in \fig{accel}(a)), due to the reconnection electric field parallel to the X-line \citep{zenitani_01,jaroschek_04,lyubarsky_liverts_08}. We find that X-point acceleration represents the first stage of energization for most of the particles that will end up with high energies. Particles pre-accelerated at X-points continue to gain energy by compression \citep{lyubarsky_liverts_08, giannios_10}, as they are advected from the X-point into the closest island. No major systematic changes of energy are observed after the particles get trapped within magnetic islands. Different X-points may provide different energy gains, as shown by the red (respectively, blue) curves in panel (a), which refer to high-energy positrons that get accelerated at the X-point above (respectively, below) the small magnetic island in the middle of the current sheet. In summary, the energy gain of a given particle is determined on the one hand by the strength of the reconnection electric field at the X-point, and on the other hand by the time available for acceleration, as the particle flows from the X-point into the closest island.

In retrospect, the importance of X-points for the generation of nonthermal particles is not surprising. In magnetically-dominated flows, particles can untie from magnetic field lines and be injected into the acceleration process only in regions where the electric field is comparable or exceeds the magnetic field, i.e., green or blue regions in the current sheets of \fig{fluidsh}(c) (where we plot the difference $\epsilon_{B}-\epsilon_{E}$ between magnetic and electric energy). Based on this criterion, X-points are natural sites of particle energization, since the magnetic field vanishes there. As discussed above, the main agent of particle acceleration at X-points is the reconnection electric field, which for the magnetic geometry employed in this work is oriented along $z$ (more precisely, along $+\bmath{\hat{z}}$ if the magnetic field $B_y$ is negative to the left of the current sheet and positive to the right; along $-\bmath{\hat{z}}$ in the opposite case). The sign of the reconnection electric field leaves an imprint on the particle $z$-momentum. As shown by the positron transverse phase space in \fig{fluidtot}(f), current sheets where the reconnection field is expected to be along $+\bmath{\hat{z}}$ are mostly populated by high-energy positrons with $\gamma\beta_{\rm z}>0$ (and electrons with $\gamma\beta_{\rm z}<0$), and the opposite holds where the reconnection field is along $-\bmath{\hat{z}}$.

Accelerated by the reconnection electric field, high-energy particles stream at first along the $z$ direction, and then, by effect of the reconnected magnetic field $B_x$, they are advected along $y$ into the closest island. In our 2D simulations, where the $z$ dimension is not resolved, the relevant timescale for acceleration is the time to drift from an X-point into the closest magnetic island. As the flow propagates downstream from the fast shock, magnetic islands grow and merge  (see  \fig{fluidsh}(a)), so that the typical distance between an X-point and the closest island increases, allowing particles to be accelerated to higher energies. This explains why the upper cutoff in the particle spectra of \fig{fluidsh}(f) shifts to higher energies for current sheets farther from the fast shock.\footnote{Another factor that drives the trend in the upper spectral cutoff of \fig{fluidsh}(f), though less important than the timescale argument mentioned above, is the increase in the strength of the reconnection field as the flow is compressed toward the hydrodynamic shock.}

Since our 2D simulations cannot resolve potential perturbations to the particle motion along $z$, the validity of our conclusions may be questioned, if the structure of X-points in a realistic 3D scenario is much different from the 2D picture analyzed here. In particular, in 3D we expect the current sheet to be folded, by effect of the so-called drift-kink instability \citep{zenitani_05,zenitani_07}. The characteristic wavelength of the kink mode in the $xz$ plane will introduce a different length scale, that could potentially compete with the distance between X-points and islands (in the $xy$ plane) in constraining the maximum energy of accelerated particles.  In \S\ref{app:3d} we address this important point, and show that the latter constraint is more restrictive than the former, at least for the parameters explored in this work. It follows that our 2D simulations with in-plane fields can capture very well the main 3D properties of the shock, and in particular the physics of particle acceleration.

Although we find that the reconnection electric field at X-points is the main agent of particle acceleration, we have verified, by tracing the orbits of a large number of particles extracted directly from the simulation, that a variety of other mechanisms can play a role in particle energization. During the coalescence of magnetic islands, particles may be accelerated by the anti-reconnection electric field at the X-point located in between the two merging islands \citep{jaroschek_04, oka_10}. Alternatively, particle energization may be due to a kind of first-order Fermi process within magnetic islands, when particles reflect from the ends of contracting islands following coalescence \citep{drake_06}. Also, the second-order Fermi process can operate between randomly-moving islands just downstream from the hydrodynamic shock. The two latter mechanisms change primarily the components of particle momentum along the plane where the magnetic field lies (the $xy$ plane, in our simulations). On the contrary, for reconnection and anti-reconnection at X-points, it is the momentum component along the reconnection electric field (along $z$, in our case) that receives the greatest boost. 

The energization mechanisms discussed so far are peculiar to the process of magnetic reconnection, being seeded by the field dissipation itself, or by the resulting fluid motions. In shock-dominated systems, two more acceleration processes can be relevant, whose free energy comes from the converging motion of the pre-shock and post-shock flows. In first-order Fermi acceleration (or diffusive shock acceleration, DSA), particles diffuse back and forth across the shock front and gain energy by scattering from magnetic turbulence embedded in the converging flows \citep[e.g.,][]{blandford_ostriker_78,bell_78,blandford_eichler_87}. In magnetized flows, particles may also gain energy directly from the large-scale motional electric field while they gyrate around the shock front, a process known as shock-drift acceleration \citep[SDA; e.g.,][]{chen_75,webb_83,begelman_kirk_90}. As we describe in the following section, the SDA process is most efficient in shocks where $\langle B_y\rangle_\lambda/B_0\gtrsim0.015$, whereas the DSA mechanism can only operate if the stripe-averaged field is small, $\langle B_y\rangle_\lambda/B_0\lesssim0.015$.

\section{Dependence on the Wind Properties}\label{sec:cond}
In this section, we investigate how the physical conditions in the wind can affect the structure of the shock and the physics of magnetic reconnection, with particular focus on the particle spectrum downstream from the hydrodynamic shock. We will explore the dependence of our results on the stripe wavelength (in \S\ref{sec:lambda}), the wind magnetization (in \S\ref{sec:sigma}), the strength of the stripe-averaged field (in \S\ref{sec:alpha}), and the wind bulk Lorentz factor (in \S\ref{sec:gamma0}).

The 1D model presented by \citet{petri_lyubarsky_07} predicts that magnetic dissipation should not appreciably affect the dynamics of the flow, provided that $\lambda_{\comp}/\sigma\gtrsim8\,\xi_1/5$. Here, $\lambda_{\comp}=\lambda/(\comp)$ is the stripe wavelength in units of the relativistic skin depth in the cold wind, and $\xi_1=6-10$ is the ratio of downstream sheet thickness to downstream Larmor radius (its value is calibrated with 1D PIC simulations). In this case, the striped structure of the wind will be preserved downstream, just compressed, and the flow will satisfy the MHD jump conditions for a strongly magnetized fluid. On the contrary, full dissipation of the alternating fields should occur if $\lambda_{\comp}/\sqrt{\sigma}\lesssim4\,\xi_1/5$, when the thickness of downstream current sheets becomes comparable to the distance between neighboring sheets. In this case, \citet{petri_lyubarsky_07} find that the post-shock flow is purely hydrodynamical, since all the magnetic field energy has been transferred to the particles.

As we discuss below, the results of our 2D and 3D simulations suggest a different picture. We find that complete dissipation of the alternating fields  by magnetic reconnection is a common by-product of the shock evolution, even in the cases when the 1D model of \citet{petri_lyubarsky_07} predicts that the stripes should be preserved downstream from the shock. As suggested by \fig{fluidsh}, multi-dimensional simulations are of key importance to correctly capture the evolution of the tearing-mode instability, which is ultimately responsible for field dissipation and particle acceleration.

Even though complete field dissipation (and transfer of field energy to the particles) occurs irrespective of the wind properties, the shape of the downstream particle spectrum is sensitive to the stripe wavelength and the wind magnetization. In most cases, the particle spectrum can be described as a power law in energy, with a flat slope of index $1<p<2$. The slope is moderately sensitive to the pre-shock parameters, but it is the width of the spectrum that depends most dramatically on the properties of the wind. If the spectrum can be modeled as a power law of index $p$ between $\gamma_{\rm min}$ and $\gamma_{\rm max}$, we can derive a relation between the lower and upper cutoffs by requiring that the mean particle Lorentz factor  be $\langle\gamma\rangle=\gamma_0(\sigma+1)$, as expected for full dissipation of magnetic energy. We obtain 
\be\label{eq:gmax}
\left(\frac{\gamma_{{\rm min}}}{\gamma_0}\right)^{p-1} \left(\frac{\gamma_{{\rm max}}}{\gamma_0}\right)^{2-p}=\frac{2-p}{p-1}\,(\sigma+1)~~.
\ee
For $1<p<2$ and fixed $\sigma$, we see that this expression includes both the case of  narrow distributions with $\gamma_{\rm min}\simeq\gamma_{\rm max}\simeq\gamma_0\sigma$, and the case of wide spectra with $\gamma_{\rm min}\simeq\gamma_0$ and  $\gamma_{\rm max}\simeq\gamma_0\sigma^{1/(2-p)}$. As we show in \S\ref{sec:lambda}, the stripe wavelength is the main factor, at fixed wind magnetization, that controls the transition from narrow Maxwellian-like spectra to broad power-law tails.

In Figs.~\fign{speclam}-\fign{scalesig}, \fign{specdc} and \fign{specgam}, we show the dependence of the downstream particle spectrum on the conditions in the wind. In most cases, the spectrum is measured between $700\comp$ and $300\comp$ downstream from the hydrodynamic shock.\footnote{The width of the slab where we compute the spectrum is calibrated such that to average out all potential variations along $x$, and to provide an estimate of the characteristic downstream spectrum.} In the subpanel of each figure, we estimate the mean particle kinetic energy $\langle\gamma\rangle/\gamma_0$ (black line) and the width of the particle spectrum, measured as the ratio $\gamma_{\rm max}/\gamma_{\rm min}$ of upper to lower energy cutoff (red line). More precisely, $\gamma_{\rm min}$ will be defined as the Lorentz factor where most of the particles lie (i.e., where $\gamma \,dN/d\gamma$ peaks), whereas $\gamma_{\rm max}$ will correspond to the location where most of the energy resides (i.e., where $\gamma^2 dN/d\gamma$ is at maximum).

\subsection{Dependence on the Stripe Wavelength}\label{sec:lambda}
\begin{figure}[tbp]
\begin{center}
\includegraphics[width=0.5\textwidth]{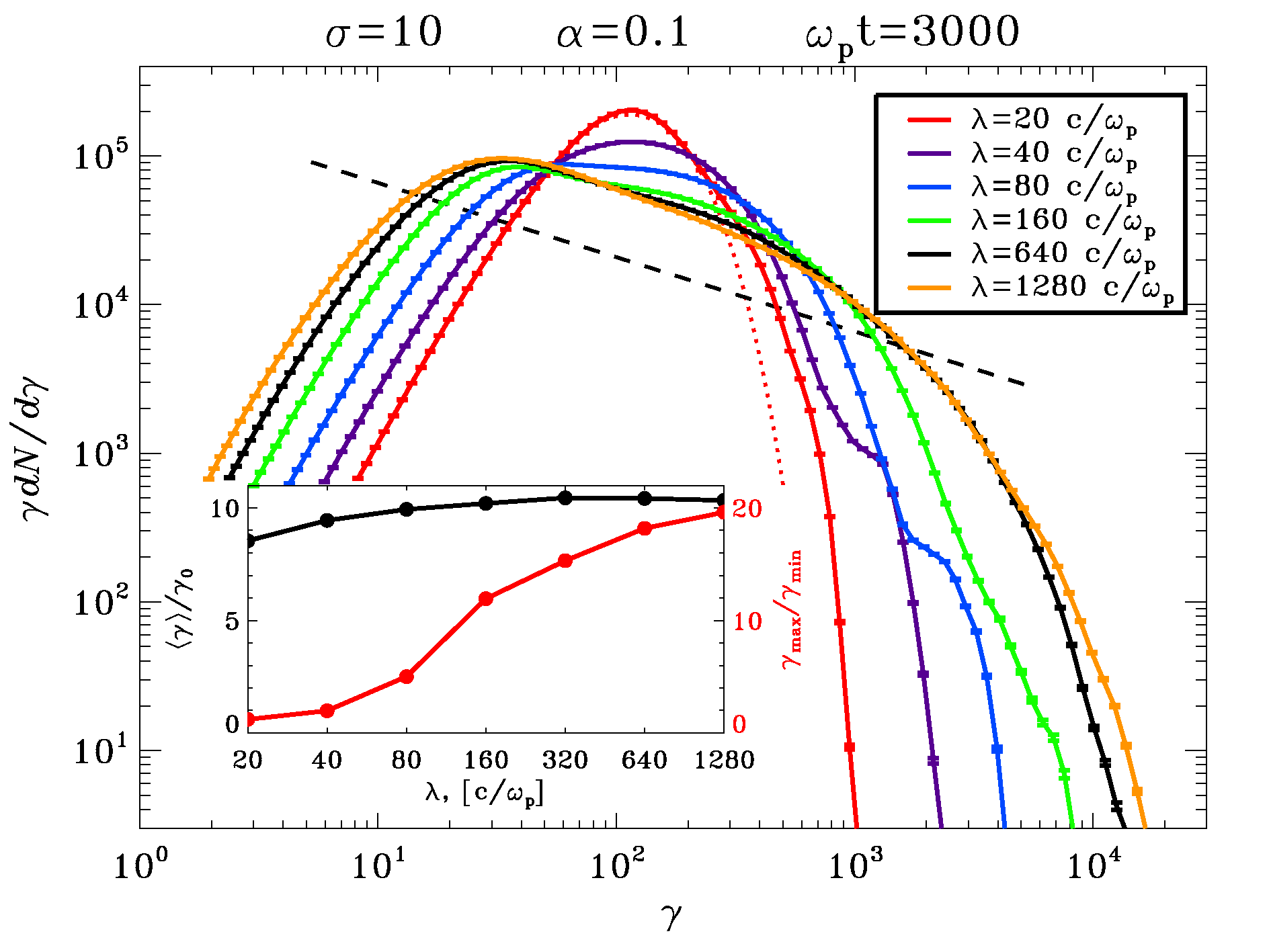}
\caption{Downstream particle spectrum at $\ompt=3000$ for different stripe wavelengths $\lambda$, in a flow with $\sigma=10$ and $\alpha=0.1$. We vary the stripe wavelength from $\lambda=20\comp$ up to $\lambda=1280\comp$. The dotted  red line is a Maxwellian with the same average energy as the spectrum colored in red (which refers to $\lambda=20\comp$); the dashed black line indicates a power-law distribution with slope $p=1.5$. In the subpanel, the black line shows the average Lorentz factor of downstream particles $\langle\gamma\rangle$ as a function of the stripe wavelength $\lambda$ (axis on the left side). In the same subpanel, the red line presents the dependence of $\gamma_{\rm max}/\gamma_{\rm min}$ on $\lambda$ (axis on the right side). Here, $\gamma_{\rm min}$ is defined as the location where $\gamma dN/d\gamma$ peaks (i.e., where most of the particles reside), whereas  $\gamma_{\rm max}$ is the Lorentz factor where $\gamma^2 dN/d\gamma$ peaks (i.e., where most of the energy lies).}
\label{fig:speclam}
\end{center}
\end{figure}

\fig{speclam} shows how the downstream particle spectrum changes for different stripe wavelengths $\lambda$, keeping fixed the magnetization $\sigma=10$, the stripe-averaged field $\langle B_y\rangle_\lambda/B_0\simeq0.05$ (corresponding to $\alpha=0.1$), and the bulk Lorentz factor $\gamma_0=15$. We explore a wide range of wavelengths, from $\lambda=20\comp$ up to $\lambda=1280\comp$.

Regardless of the value of $\lambda$, the structure of the flow resembles the picture presented in \S\ref{sec:shock}. A fast MHD shock propagates into the striped wind, compressing the incoming current sheets and triggering magnetic reconnection. Due to reconnection, the striped structure is eventually erased, and a hydrodynamic shock forms. Downstream from the shock, the flow is nearly unmagnetized, apart from the field component associated with the shock-compression of the stripe-averaged field $\langle  B_y\rangle_\lambda$. It follows that the downstream mean kinetic energy per particle should be close to the value $\langle\gamma\rangle/\gamma_0=\sigma+1$ expected in the case of complete field dissipation. The value of $\langle\gamma\rangle$ measured in our simulations (black line in the subpanel of \fig{speclam}) is approximately constant with respect to $\lambda$, and is consistent with such expectation.\footnote{The trend of smaller values of $\langle\gamma\rangle/\gamma_0$ with decreasing $\lambda$ (black line in the subpanel of \fig{speclam}) is due to the fact that the fractional contribution of current sheets (with fixed thickness $2\Delta\simeq2\comp$) to the upstream flow is larger for smaller $\lambda$. When averaged over one stripe wavelength, this results in a lower upstream energy per particle for smaller $\lambda$ (including both magnetic and kinetic contributions), which is then reflected in the post-shock value of $\langle\gamma\rangle/\gamma_0$.} 
Clearly, this is in contrast with the 1D model of \citet{petri_lyubarsky_07}, which would predict negligible field dissipation if $\lambda\gtrsim130\comp$ (for $\xi_1=8$ and $\sigma=10$).

Even though the mean kinetic energy per particle does not appreciably vary with $\lambda$, the shape of the spectrum does change, with a clear tendency for broader spectra at longer stripe wavelengths. As shown by the red line in the subpanel of \fig{speclam}, the ratio $\gamma_{\rm max}/\gamma_{\rm min}$ increases with wavelength from  $\gamma_{\rm max}/\gamma_{\rm min}\simeq1$ (at $\lambda=20\comp$) up to $\gamma_{\rm max}/\gamma_{\rm min}\simeq20$ (at $\lambda=1280\comp$). Longer wavelengths yield smaller values of $\gamma_{\rm min}$ (see main plot), which then imply larger values of $\gamma_{\rm max}$, given the relation in \eq{gmax}. 

This trend can be easily understood by considering the structure of the flow for different stripe wavelengths. As described in \S\ref{sec:shock}, the hydrodynamic shock is located at the point where reconnection islands grow large enough to fill the entire space in between neighboring current sheets. Since the distance between sheets is proportional to $\lambda$, the same scaling should hold for the size of reconnection islands, just upstream of the hydrodynamic shock.\footnote{In the remaining of this section, by ``reconnection islands''  we will be referring only to the islands in the current sheet just ahead of the hydrodynamic shock, where the shape of the downstream particle spectrum is established, as demonstrated by \fig{fluidsh}(f).} Shorter wavelengths will then result in more numerous islands of smaller size, whereas fewer but bigger islands will be present for longer $\lambda$ (see \fig{fluidsh}). Since an X-point exists in between each pair of neighboring islands (belonging to the same current sheet), the number of X-points per unit length (along the current sheet) will be larger for smaller $\lambda$. For short wavelengths, most of the incoming particles will likely pass in the vicinity of one of the numerous X-points, as the flow crosses the hydrodynamic shock. The energy evolution of one particle will then be similar to that of any other particle, with its Lorentz factor increasing from $\gamma\simeq\gamma_0$ up to $\gamma\simeq\gamma_0(\sigma+1)$, as the particle gains energy from the annihilating fields. This explains why for short stripe wavelengths ($\lambda=20\comp$, red curve in \fig{speclam}) the particle spectrum is so narrow, and similar to a Maxwellian distribution (red dotted curve). In summary, for short stripe wavelengths, all particles are equally close to an X-point, so they gain comparable amounts of energy.

On the other hand, for long wavelengths, the energy evolution of different particles can be extremely diverse, as discussed in \S\ref{sec:accel}. Particles that stay away from X-points are likely to remain cold, and retain the energy $\gamma\simeq\gamma_0$ they started with. On the contrary, particles that pass through an X-point can gain a significant amount of energy. With increasing stripe wavelength, more particles will belong to the former group, and fewer to the latter, since the number of X-points decreases. This explains why the peaks in the spectra of \fig{speclam}, which track the downstream Lorentz factor of the ``typical'' particle, shift to lower energies with increasing stripe wavelength. For the same reason, the best-fitting power law becomes softer, with spectral slope increasing from $p\simeq1.0$ to $p\simeq1.5$ (indicated as a black dashed line in \fig{speclam}). In response to a lower $\gamma_{\rm min}$ and a steeper slope, the upper cutoff of the spectrum shifts to higher energies, as predicted by \eq{gmax}. This is just a consequence of the longer time available for acceleration, at larger stripe wavelengths. As discussed in \S\ref{sec:accel}, particles injected at X-points are continuously accelerated by the reconnection electric field, while they drift from the X-point into the closest island. Since the spacing between neighboring islands (in the same current sheet) increases with stripe wavelength, the injected particles will be able to stay in the acceleration region for longer times (and so, to reach higher energies), if $\lambda$ is larger.

As we discuss in \S\ref{sec:sigma}, the threshold between short and long wavelengths depends on the wind magnetization. Here, by ``short $\lambda$'' we generically mean all cases in which the spectrum is narrow, similar to a Maxwellian. In contrast, the cases with ``long $\lambda$'' have broad power-law spectra with slope $1<p<2$. We find that the condition proposed for full dissipation by \citet{petri_lyubarsky_07}, namely that $\lambda_{\comp}/\sqrt{\sigma}\lesssim4\,\xi_1/5$, works fairly well to discriminate between short and long wavelengths. For $\xi_1=8$, such criterion would predict Maxwellian-like spectra for $\lambda\lesssim20\comp$ (if $\sigma=10$), $\lambda\lesssim45\comp$ (if $\sigma=50$), and  $\lambda\lesssim65\comp$ (if $\sigma=100$), which is indeed what we observe in our simulations. In retrospect, this is not surprising. In fact, the condition $\lambda_{\comp}/\sqrt{\sigma}\lesssim4\,\xi_1/5$ comes from requiring that the downstream sheet thickness be larger than the separation of neighboring sheets. In this limit, no distinction should persist between the plasma within (versus outside) the current sheets, i.e., all particles should share the same energy evolution. This obviously results in a Maxwellian distribution.

Finally, we point out that the minor bump emerging at high energies in the spectra of \fig{speclam} (most notably for wavelengths $\lambda\lesssim160\comp$) is populated by particles that are energized via the shock-drift mechanism (SDA) at the hydrodynamic shock. Such particles gain energy from the stripe-averaged motional electric field $\langle  E_z\rangle_\lambda\simeq\langle  B_y\rangle_\lambda$, while they gyrate around the shock. Since $\langle  E_z\rangle_\lambda\geq0$ for our simulation setup, shock-drift accelerated positrons should preferentially have $\gamma\beta_{\rm z}>0$, whereas the opposite should hold for electrons. The few high-$\gamma\beta_{\rm z}$ positrons seen at $x\simeq1100\comp$ in panel (f) of \fig{fluidtot} are indeed accelerated via the SDA mechanism. They gyrate around the shock a few times, before being advected downstream by the stripe-averaged magnetic field. Given the limited number of acceleration cycles, the SDA component in the spectra of \fig{speclam} does not extend in time to higher energies. So, SDA is not a promising candidate to produce broad power-law tails.

The efficiency of injection into the SDA process is higher for shorter stripe wavelengths, as \fig{speclam} suggests. As seen from the pre-shock frame, a particle coming from the downstream side can be accelerated via SDA in the upstream stripe-averaged field only if it can sample the full wavelength of the striped wind during its upstream residence time. Since particles returning upstream are caught up by the shock after having travelled a distance $r_{L,\rm u}/\gamma_0$ (here, $r_{L,\rm u}$ is the particle Larmor radius measured in the upstream frame), we require $\gamma_0(\lambda/2)\lesssim r_{L,\rm u}/\gamma_0$ for efficient SDA. If $\gamma= \xi_2\langle\gamma\rangle\simeq \xi_2\gamma_0\sigma$ is the  Lorentz factor of particles at the high-energy end of the distribution ($\xi_2=1-10$, depending on $\lambda$), the previous condition can be rewritten as $\lambda_{\comp}/\sqrt{\sigma}\lesssim 2\,\xi_2$. For fixed magnetization, the criterion for efficient SDA is more easily satisfied at short stripe wavelengths, which explains why the normalization of the high-energy bump in \fig{speclam} increases for smaller $\lambda$. 

We remark that, apart from the factor $\xi_2$, which  could in principle depend (though weakly) on $\lambda$ and $\sigma$, the criterion for efficient SDA involves the same combination of wind magnetization and stripe wavelength (namely, $\lambda_{\comp}/\sqrt{\sigma}$) as the condition discussed at the beginning of this section, which regulates how closely the particle spectrum resembles a Maxwellian distribution \citep[see also][]{petri_lyubarsky_07}. Since $\lambda_{\comp}$ is the wind wavelength normalized to the skin depth of the \tit{pre-shock} fluid, and since the average particle energy increases by a factor of  $\simeq\sigma$ across the shock as a result of field dissipation, then $\lambda_{\comp}/\sqrt{\sigma}$ is just the wind wavelength measured in units of the \tit{post-shock} plasma skin depth (apart from factors of order unity). In \S\ref{sec:sigma}, we further comment on the importance on the parameter $\lambda_{\comp}/\sqrt{\sigma}$ in regulating the physics of the shock and the shape of the downstream particle spectrum.

\subsection{Dependence on the Wind Magnetization}\label{sec:sigma}
\begin{figure}[tbp]
\begin{center}
\includegraphics[width=0.5\textwidth]{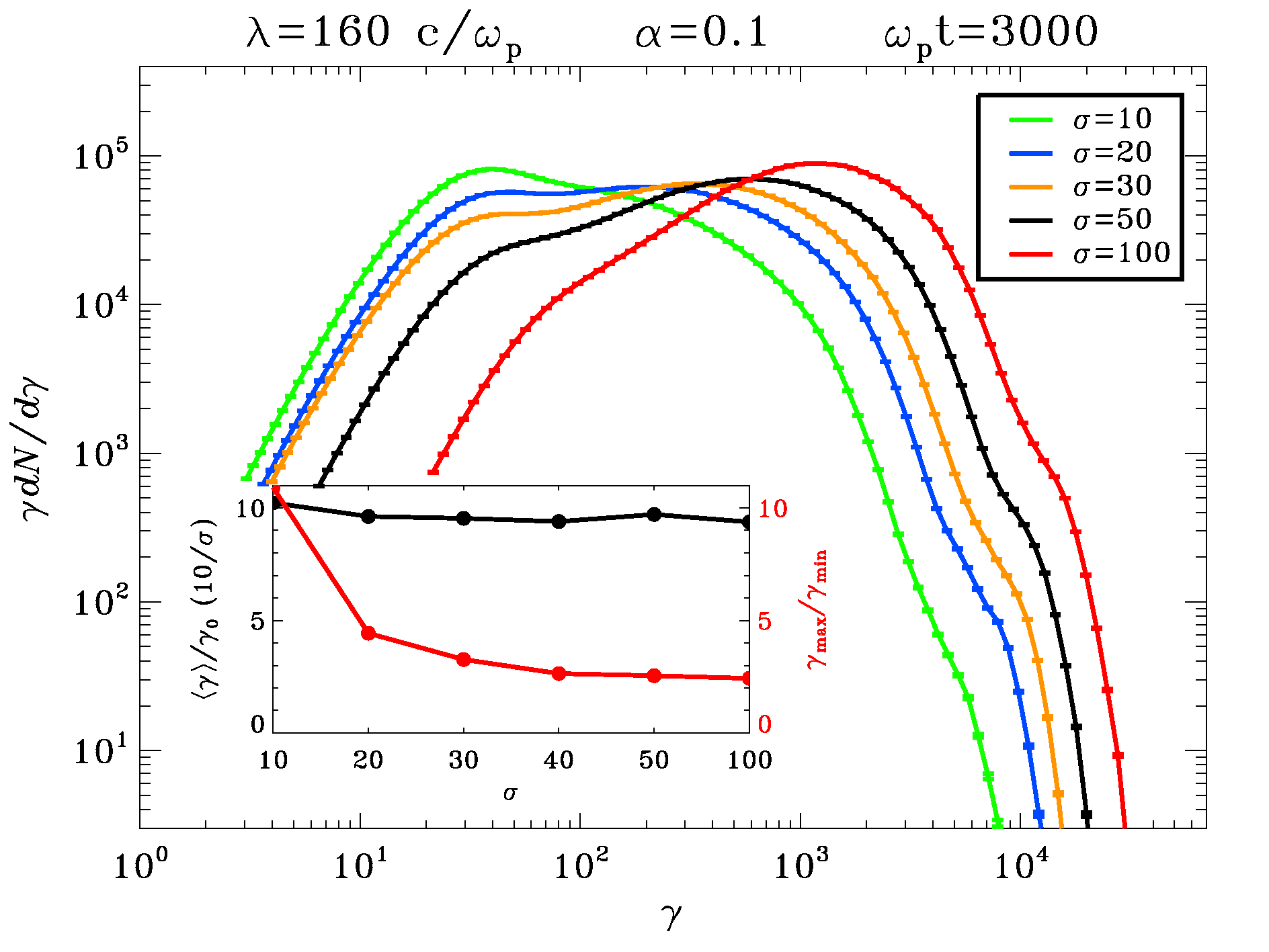}
\caption{Downstream particle spectrum at $\ompt=3000$ for different magnetizations $\sigma$, in a flow with $\lambda=160\comp$ and $\alpha=0.1$. We vary the wind magnetization from $\sigma=10$ to $\sigma=100$. In the subpanel, the black line shows the average Lorentz factor $\langle \gamma \rangle$ of downstream particles as a function of $\sigma$ (axis on the left side). The value of $\langle \gamma \rangle$ is normalized to the total  energy per particle (kinetic + electromagnetic) in the pre-shock flow, i.e., $\gamma_0\sigma$. In the same subpanel, the red line presents the dependence of the spectral width $\gamma_{\rm max}/\gamma_{\rm min}$ on $\sigma$ (axis on the right side).}
\label{fig:specsig}
\end{center}
\end{figure}

In this section, we investigate the dependence of our results on the magnetization of the wind. \fig{specsig} shows how the downstream spectrum changes with $\sigma$, keeping fixed the stripe wavelength $\lambda=160\comp$, the stripe-averaged field $\langle B_y\rangle_\lambda\simeq0.05$ (corresponding to $\alpha=0.1$), and the bulk Lorentz factor $\gamma_0=15$. We explore a wide range of magnetizations, from $\sigma=10$ up to $\sigma=100$.

We find that efficient annihilation of magnetic fields by shock-driven reconnection, and transfer of magnetic energy to the particles, occur irrespective of the wind magnetization. As shown by the black line in the subpanel of \fig{specsig}, the average Lorentz factor of downstream particles approaches in all cases the value $\langle \gamma\rangle\simeq\gamma_0\sigma$ expected for complete field dissipation. As the wind magnetization increases, the spectrum shifts to higher Lorentz factors (just because $\langle \gamma\rangle\propto \sigma$) and it changes in shape, with the part at low energies  getting de-populated at the expense of the high-energy component. As a result, the particle spectrum, which could be fitted as a broad power law of index $p\simeq1.2$ for $\sigma=10$ (blue curve), approaches a Maxwellian-like  distribution for $\sigma=100$ (red curve). As a consequence, the ratio $\gamma_{\rm max}/\gamma_{\rm min}$, a proxy for the width of the spectrum, becomes smaller for increasing $\sigma$ (red line in the subpanel of \fig{specsig}, with axis on the right). 

This is the same trend observed when decreasing $\lambda$ at fixed $\sigma$, as discussed in \S\ref{sec:lambda}. In the limit of very high magnetizations, for fixed wavelength (or very short $\lambda$, for fixed $\sigma$), the downstream energy spectrum approaches a Maxwellian distribution, when $\lambda_{\comp}/\sqrt{\sigma}\lesssim4\,\xi_1/5$ (here, $\xi_1=6-10$). For $\lambda=20\comp$, $40\comp$, and $80\comp$, we have verified that the value of $\sigma$ above which the spectrum resembles a Maxwellian is in good agreement with this criterion. For $\lambda=160\comp$, we would expect a Maxwellian distribution only for $\sigma\gtrsim300$ (if $\xi_1=8$), beyond the range of magnetizations we explore. However, the trend toward a single-component distribution (as $\sigma$ increases) is already clear within the limited range of magnetizations investigated in \fig{specsig}.

The dominance of the high-energy component  at large magnetizations can be understood in terms of the basic properties of the flow. In experiments with fixed $\sigma$ and different $\lambda$ (see \S\ref{sec:lambda}), we have related the shape of the spectrum to the size of magnetic islands just ahead of the hydrodynamic shock, that scales with stripe wavelength. Here $\lambda$ is fixed, and the change in spectral shape with magnetization should instead be attributed to the different reconnection rate, namely the velocity $\beta_{\rm rec}$ at which cold plasma flows into the X-points. \citet{zenitani_hesse_09} have measured the reconnection rate in high-$\sigma$ pair plasmas via 2D two-fluid MHD simulations of undriven reconnection, finding that the inflow velocity scales roughly as $\beta_{\rm rec}\simeq0.07\sqrt{\sigma}$ in the range $\sigma=10-100$. We have confirmed this scaling in our simulations, by measuring as a function of magnetization (at fixed $\lambda$) the distance between the fast and the hydrodynamic shock. Everything else being fixed, this distance should be inversely proportional to the reconnection rate, and in the range of magnetizations explored in our study we find that the reconnection speed scales as $\beta_{\rm rec}\propto\sqrt{\sigma}$.

As the reconnection rate increases with magnetization, a larger number of particles can be advected into a given X-point per unit time, which results in more particles experiencing the reconnection electric field. In other words, at larger $\sigma$ it is more likely for the ``typical'' particle to enter an X-point and gain energy from field dissipation. This explains why for high magnetizations the spectra in \fig{specsig} are dominated by a single component of hot particles accelerated by the reconnection field. A similar argument explains why, for large values of $\sigma$, the spectrum does not extend much beyond the average Lorentz factor $\langle\gamma\rangle$. In response to the increase of inflow speed with $\sigma$, also the outflow velocity from a given X-point into the closest island increases with magnetization. For fixed $\lambda$, which corresponds to a fixed spacing between X-points and magnetic islands, the time available for acceleration by the reconnection field will be shorter, for larger $\sigma$. So, the maximum energy at which particles can be accelerated will be lower (relative to the average Lorentz factor $\langle\gamma\rangle$), for higher magnetizations.

Finally, we notice that the minor bump emerging at high energies from the spectra of \fig{specsig}, which is populated by particles accelerated via the SDA mechanism (see \S\ref{sec:lambda}), becomes more prominent for larger magnetizations. This is in good agreement with the criterion $\lambda_{\comp}/\sqrt{\sigma}\lesssim 2\,\xi_2$ discussed at the end of \S\ref{sec:lambda}, which dictates that injection into the SDA process should be easier for higher $\sigma$ (at fixed $\lambda$).

In summary, we find that both the physics of shock-driven reconnection, which ultimately governs the shape of the particle spectrum, and the efficiency of SDA are controlled by the same parameter $\lambda_{\comp}/\sqrt{\sigma}$, namely the stripe wavelength measured in units of the  \tit{post-shock} skin depth ($\simeq\sqrt{\sigma}\comp$). We expect that the main properties of the shock, and in particular the high-energy end of the downstream particle spectrum, should be relatively insensitive to variations in $\lambda$ or $\sigma$, provided that the ratio $\lambda_{\comp}/\sqrt{\sigma}$ is kept constant.

\begin{figure}[tbp]
\begin{center}
\includegraphics[width=0.5\textwidth]{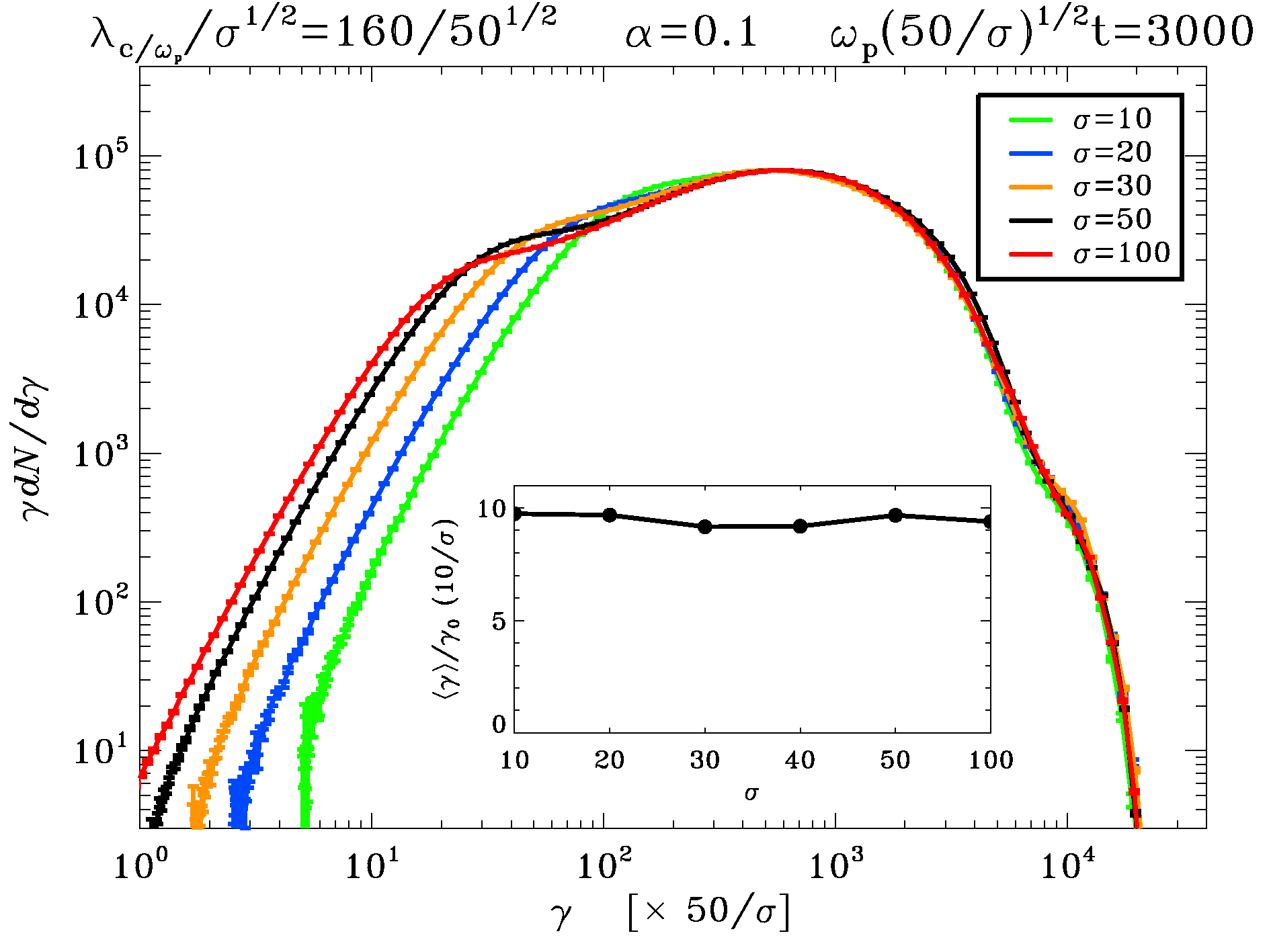}
\caption{Downstream particle spectrum for different values of $\lambda$ and $\sigma$, but such that the ratio $\lambda_{\comp}/\sqrt{\sigma}=160/\sqrt{50}\simeq22.5$ is fixed (as well as $\alpha=0.1$). Here, $\lambda_{\comp}$ is the wind wavelength in units of the skin depth in the \tit{pre-shock} fluid, so $\lambda_{\comp}/\sqrt{\sigma}$ is roughly the wavelength normalized to the skin depth of the \tit{post-shock} fluid. The comparison is performed at the same time, in units of the relativistic plasma frequency of the  \tit{post-shock}  flow ($\simeq\omega_{\rm p}/\sqrt{\sigma}$). Spectra are shifted along the $x$-axis by $50/\sigma$ to facilitate comparison with the case $\sigma=50$. The black line in the subpanel shows the average downstream Lorentz factor, as in \fig{specsig}.}
\label{fig:scalesig}
\end{center}
\end{figure}

In \fig{scalesig} we investigate this prediction by comparing the downstream spectrum among flows that have the same value of $\lambda_{\comp}/\sqrt{\sigma}=160/\sqrt{50}\simeq22.5$. We vary the wind magnetization from $\sigma=10$ up to $\sigma=100$, or equivalently the stripe wavelength from $\lambda=72\comp$ up to $\lambda=226\comp$. The spectra in \fig{scalesig} are shifted along the $x$-axis by $50/\sigma$, to facilitate comparison with the reference case $\sigma=50$ (black curve). This compensates for the overall shift to higher Lorentz factors due to the increase in $\langle\gamma\rangle\propto\sigma$, as  expected  for complete field dissipation.\footnote{Strictly speaking, the comparison in \fig{scalesig} is performed at the same $\ompt/\sqrt{\sigma}$, namely at the same time in units of the \tit{post-shock}  plasma frequency ($\simeq\omega_{\rm p}/\sqrt{\sigma}$). However, we stress once again that the downstream spectrum does not appreciably evolve in time.} We confirm that the shape of the spectrum, at the high-energy end, is controlled by the single parameter $\lambda_{\comp}/\sqrt{\sigma}$. For a fixed value of this particular combination of $\lambda$ and $\sigma$, both the high-energy component of the spectrum and the bump of shock-drift accelerated particles are insensitive to variations in either $\lambda$ or $\sigma$. 

The shift of the low-energy part to higher Lorentz factors, as $\sigma$ decreases, is just an artificial consequence of our re-scaling in the $x$-axis of \fig{scalesig}. In reality, the low-energy cutoff stays fixed at some multiple of the pre-shock bulk Lorentz factor $\gamma_0$, irrespective of $\lambda$ or $\sigma$, provided $\lambda_{\comp}/\sqrt{\sigma}$ is kept constant. With decreasing $\lambda$, the number of X-points per unit length increases as $\propto\lambda^{-1}$, which would make it easier for the ``tyipical'' particle to gain energy from the dissipating fields (everything else being equal). However, for lower $\sigma$, the inflow velocity into a given X-point is also reduced, as $\propto\sqrt{\sigma}$. Evidently, for constant $\lambda_{\comp}/\sqrt{\sigma}$, the two opposite effects compensate, resulting in a fixed low-energy spectral cutoff.

\subsection{Dependence on the Stripe-Averaged Field}\label{sec:alpha}
In this section, we explore the dependence of the shock  properties on the strength of the stripe-averaged field $\langle B_y\rangle_\lambda$. The magnitude $B_0$ of the field in the cold wind is fixed by the value of the magnetization $\sigma$. A non-zero stripe-averaged field is established if the portion of each stripe with $B_y=+B_0$ has a different width than the region with $B_y=-B_0$. If $\lambda_+$ is the length of the former and $\lambda_{-}$ of the latter (with the constraint $\lambda_++\lambda_-=\lambda$), then we have $\langle B_y\rangle_\lambda/B_0=(\lambda_+-\lambda_-)/\lambda$. Alternatively, by defining $\alpha\equiv(\lambda_+-\lambda_-)/\lambda_+$, we have $\langle B_y\rangle_\lambda/B_0=\alpha/(2-|\alpha|)$. In pulsar winds, $|\alpha|\rightarrow1$ at latitudes approaching the inclination angle between magnetic and rotational axes of the pulsar (see the upper panel of \fig{simplane}). The special case $\alpha=0$, appropriate for the wind midplane, will be discussed in more detail in \S\ref{sec:zero} below.

\begin{figure}[tbp]
\begin{center}
\includegraphics[width=0.5\textwidth]{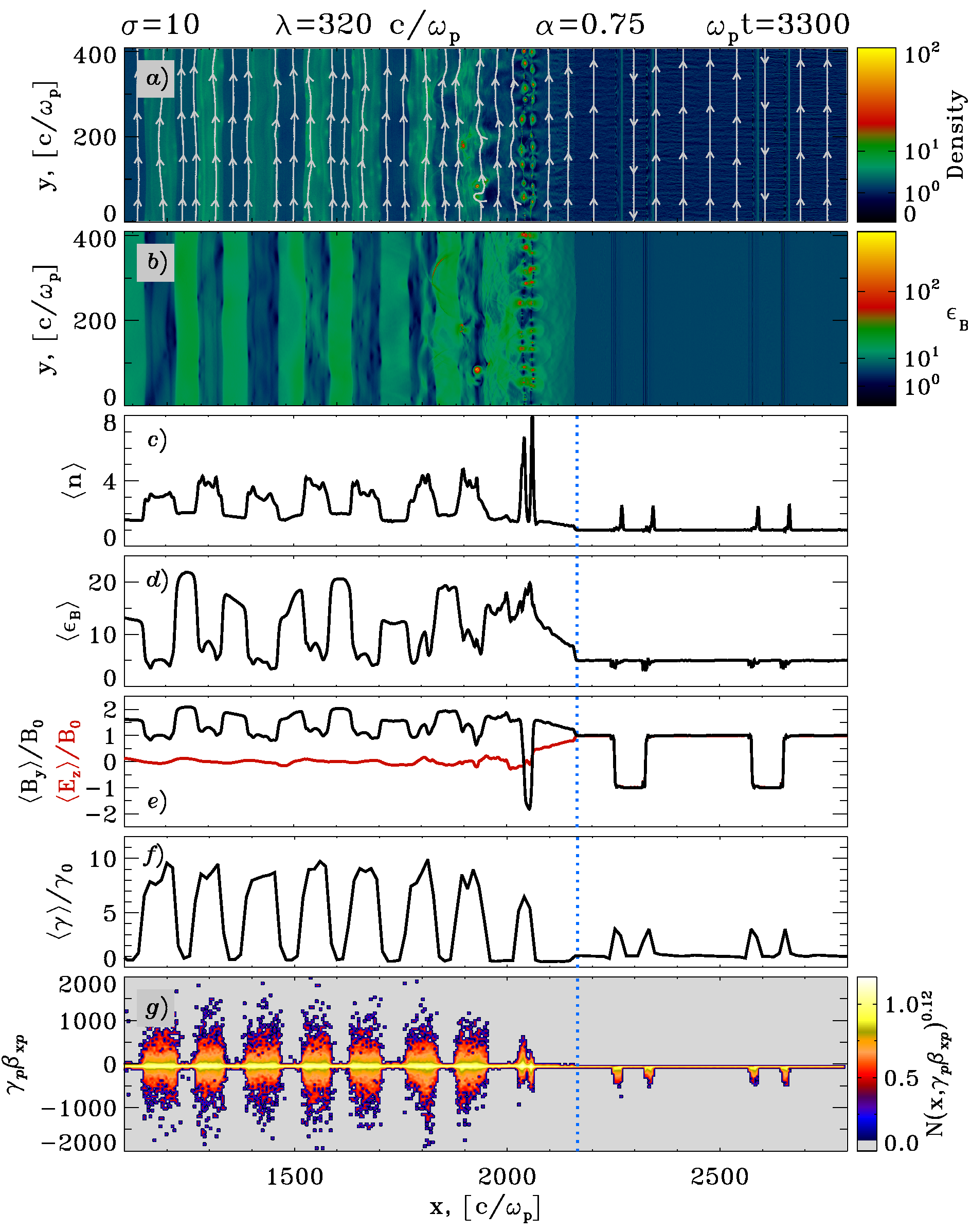}
\caption{Internal structure of the flow at $\ompt=3300$, for stripe wavelength $\lambda=320\comp$ and magnetization $\sigma=10$, zooming in on a region around the shock. The stripe-averaged field is $\langle B_y\rangle_\lambda/B_0\simeq0.6$, corresponding to $\alpha=0.75$. The hydrodynamic shock is located at $x\simeq2050\comp$, and the location of the fast MHD shock ($x\simeq2170\comp$) is indicated with a vertical dotted blue line. The following quantities are plotted: (a) 2D plot in the simulation plane of the particle number density, in units of the upstream value, with contours showing the magnetic field lines; (b) 2D plot of the magnetic energy fraction $\epsilon_B\equiv B^2/8 \pi \gamma_0 m n_{c0} c^2$; (c) $y$-averaged particle number density; (d) $y$-averaged magnetic energy fraction; (e) $y$-averaged magnetic field $B_y$ (black line) and electric field $E_z$ (red line), normalized to the upstream magnetic field $B_{0}$; (f) mean kinetic energy per particle, in units of the bulk energy at injection; (g) $x-\gb{x}$ positron phase space.}
\label{fig:fluidhigh}
\end{center}
\end{figure}

The results presented in the previous sections referred to shocks with a moderate value of $\alpha=0.1$. As a representative case of high-$\alpha$ flows, \fig{fluidhigh} shows the shock structure for $\alpha=0.75$, corresponding to $\langle B_y\rangle_\lambda/B_0\simeq0.6$. The stripe wavelength $\lambda=320\comp$ and the wind magnetization $\sigma=10$ are chosen to facilitate comparison with the results discussed in \S\ref{sec:shock}. As in the case $\alpha=0.1$, the pre-shock flow interacts at first with a fast MHD shock (at $x\simeq2170\comp$ in \fig{fluidhigh}, indicated with a vertical dotted blue line), that triggers magnetic reconnection. With increasing $\alpha$ (and so, with increasing $\langle B_y\rangle_\lambda$), a smaller fraction of the incoming Poynting flux is in the form of alternating fields, and as such available for dissipation. For $\alpha>0$ (which implies $\lambda_+>\lambda_-$), reconnection behind the fast shock will eventually annihilate the field where $B_y=-B_0$, but a finite region  where the field stays $B_y=+B_0$ will survive. Dissipation of the alternating component is completed when the size of reconnection islands approaches  $\lambda_-$ (or, in general, the smallest between $\lambda_+$ and $\lambda_-$), as shown by the 2D plots of density and magnetic energy in \fig{fluidhigh}(a) and (b), respectively. At this point, a second shock forms (at $x\simeq2050\comp$ in \fig{fluidhigh}), which would correspond  in the regime of small $\alpha$  (i.e., $\alpha\lesssim0.1$) to the hydrodynamic shock discussed in \S\ref{sec:shock}. For larger $\alpha$, the speed of this second shock increases, from $\beta_{\rm sh}\simeq1/3$, as appropriate for an unmagnetized post-shock plasma, up to $\beta_{\rm sh}\rightarrow1$, the shock velocity in a $\sigma\gg1$ flow. In the limit $|\alpha|\rightarrow 1$, the shock that in the regime $\alpha\lesssim0.1$ was identified as ``hydrodynamic''  becomes eventually degenerate with the fast MHD shock. In the case $\alpha=0.75$ shown in \fig{fluidhigh}, the fast shock stays at a distance of $\simeq100\comp$ ahead of the main shock, and they both move at $\beta_{\rm sh}\simeq0.6$.

Downstream from the main shock ($x\lesssim2050\comp$), the residual magnetic flux that has not been annihilated on the way from the fast shock survives within stripes of high magnetic field, separated by regions of relatively low magnetic content (see the 2D  and 1D plots of magnetic energy in \fig{fluidhigh}(b) and (d), respectively; and the 1D profile of $B_y$, black line in \fig{fluidhigh}(e)). The plasma inside the high-field stripes is relatively cold, not having gained much energy from field dissipation. In contrast, the low-field regions are populated with hot particles energized by the reconnection field, as shown in the 1D profile of mean kinetic energy per particle of \fig{fluidhigh}(f) and in the longitudinal positron phase space of \fig{fluidhigh}(g). We find that the post-shock flow retains a striped structure, with magnetically-dominated regions separated by relatively wide sheets of hot plasma, for values of $|\alpha|\gtrsim0.3$ (or equivalently, for $|\langle B_y\rangle_\lambda|/B_0\gtrsim0.15$). As $|\alpha|$ increases, the longitudinal extent of the high-field stripes grows, at the expense of the regions of hot dense plasma, and magnetic pressure dominates over particle pressure in the energetics of the downstream fluid.

\begin{figure}[tbp]
\begin{center}
\includegraphics[width=0.5\textwidth]{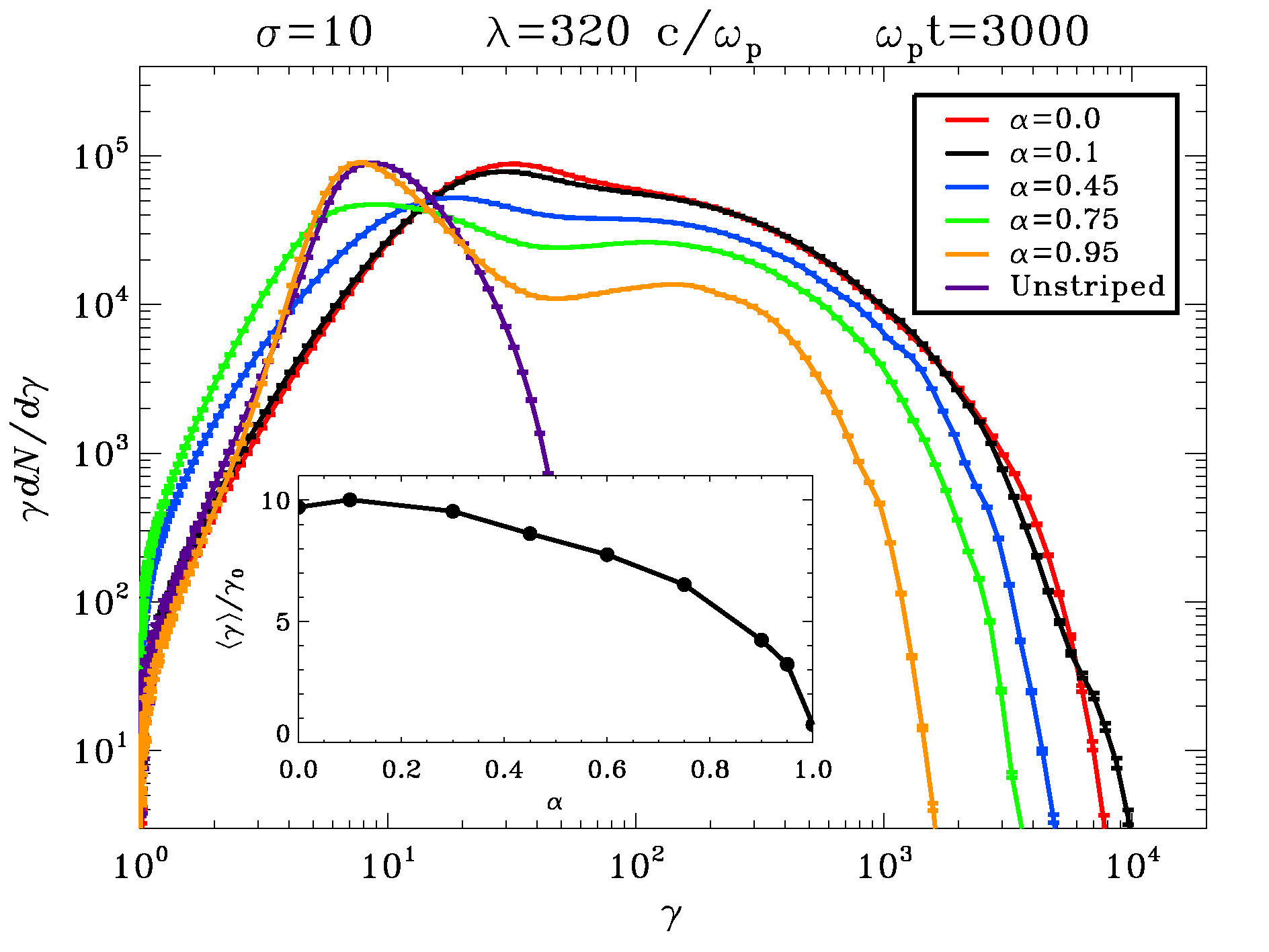}
\caption{Downstream particle spectrum at $\ompt=3000$ for different values of the stripe-averaged field $\langle B_y\rangle_\lambda$ (or equivalently, of the parameter $\alpha$), in a flow with $\lambda=320\comp$ and $\sigma=10$. We vary $\alpha$ from $\alpha=0$, when the regions with $B_y=+B_0$ and $B_y=-B_0$ have the same width $\lambda_+=\lambda_-=\lambda/2$, to the limit $\alpha\rightarrow1$,  when $\lambda_-\ll\lambda_+\simeq\lambda$. The limit of an unstriped wind is shown for reference as a purple line. The black line in the subpanel shows the average downstream Lorentz factor as a function of $\alpha$ (with $\alpha=1.0$ referring to the unstriped wind).}
\label{fig:specdc}
\end{center}
\end{figure}

The striped structure of the downstream flow leaves a clear imprint on the post-shock particle spectrum, as \fig{specdc} shows. There, we present how the particle distribution changes when the parameter $\alpha$ is varied  between $\alpha=0.0$ and $\alpha=0.95$, for fixed wavelength $\lambda=320\comp$, magnetization $\sigma=10$, and bulk Lorentz factor $\gamma_0=15$. 

For $\alpha\lesssim0.3$,  the average particle Lorentz factor downstream from the hydrodynamic shock approaches $\langle\gamma\rangle/\gamma_0\simeq\sigma+1\simeq11$, suggesting full dissipation of magnetic fields (black line in the subpanel of \fig{specdc}). The shape of the spectrum is independent of $\alpha$, in the regime $\alpha\lesssim0.1$. The spectra for $\alpha=0.0$ (red curve) and $\alpha=0.1$ (black curve) differ only at the upper end, where the  case $\alpha=0.1$ shows the high-energy bump of shock-drift accelerated particles. This component disappears for $\alpha=0.0$, when the stripe-averaged field $\langle E_z\rangle_\lambda\,=\,\beta_0\langle B_y\rangle_\lambda$ powering the SDA process vanishes.

For $\alpha\gtrsim0.3$, the post-shock particle spectrum consists of two components. The low-energy peak comes from cold plasma residing in the high-field stripes described above, whereas the high-energy part is populated by hot particles that gained energy from field dissipation. As $\alpha$ increases, the fraction of upstream Poynting flux available for dissipation decreases, which explains why the high-energy component in the spectra of \fig{specdc} gets de-populated, at the expense of the low-energy part. As a consequence, the average particle energy downstream from the shock monotonically drops from $\langle\gamma\rangle\simeq\gamma_0(\sigma+1)$, corresponding to the case of full dissipation, down to $\langle\gamma\rangle\simeq\gamma_0$, as expected for negligible field annihilation (black line in the subpanel of \fig{specdc}). In the limit $|\alpha|\rightarrow 1$ (yellow curve for $\alpha=0.95$), most of the downstream plasma is contained within the high-field regions, and the low-energy component of the spectrum approaches the result expected for an unstriped wind (purple line).\footnote{We remark that the low-energy peak in the spectrum of high-$\alpha$ flows, including the limit of unstriped wind, is not  compatible with a Maxwellian (yellow and purple lines in \fig{specdc}). Here, the particle distribution is a ring in momentum space, lying in a plane orthogonal to the field. The location of the low-energy spectral peak scales linearly with $\gamma_0$, in agreement with the jump conditions for a highly magnetized fluid.} 

With increasing $\alpha$, the high-energy component of the downstream spectrum recedes to lower energies (see the trend in \fig{specdc}, from the blue line for $\alpha=0.45$, to the yellow line for $\alpha=0.95$). As described in \S\ref{sec:lambda}, the maximum energy of particles accelerated at X-points depends primarily on the amount of time available for acceleration, which is set by the characteristic distance between a given X-point and the closest island (in the same current sheet). In turn, this scales with the maximum size of reconnection islands, which is determined by the smallest between $\lambda_+$ and $\lambda_-$ (here $\lambda_-<\lambda_+$, since $\alpha>0$). For larger $\alpha$ at fixed $\lambda$, the extent of $\lambda_-$ decreases, so the maximum energy to which particles can be accelerated by the reconnection electric field decreases. This explains the trend in the upper cutoff of the spectra in \fig{specdc}. Given that the location of the high-energy cutoff is  set by the value of $\lambda_-$ (or equivalently, by the maximum size of reconnection islands), one would expect that simulations with the same $\lambda_-$, but different choices of $\lambda$ and $\alpha>0$, should yield similar spectra, at the high-energy end. We have tested this, and we find that flows with the same $\lambda_-$ show magnetic islands of similar sizes, and their downstream spectra almost overlap at high energies. Of course, this scaling does not hold for the low-energy spectral component, since it is populated by particles that were initially in the region with $B_y=+B_0$, and are therefore insensitive to the value of $\lambda_-$.


\subsubsection{The Case of Zero Stripe-Averaged Field}\label{sec:zero}
The special case $\langle B_y\rangle_\lambda=0$  (or equivalently, $\alpha=0$), which is relevant for the equatorial plane of pulsar winds, is worth a deeper investigation. As anticipated in \S\ref{sec:lambda}, hot downstream particles can propagate back from the shock into the upstream, and cross at least one wavelength of the striped wind, provided that $\lambda_{\comp}/\sqrt{\sigma}\lesssim2\, \xi_2$, where $\xi_2=1-10$ is a numerical factor that depends weakly  on $\lambda$ and $\sigma$. For wind parameters that satisfy such criterion, the motion of hot particles moving into the upstream (the ``returning'' particles, from now on) will be regulated, to zeroth order, by the stripe-averaged fields $\langle B_y\rangle_\lambda$ and $\langle E_z\rangle_\lambda=\beta_0\langle B_y\rangle_\lambda$. For $\alpha\gtrsim0.03$ (or equivalently, $\langle B_y\rangle_\lambda/B_0\gtrsim0.015$), the returning particles will be quickly advected downstream by the stripe-averaged fields, after being accelerated via the SDA mechanism discussed in \S\ref{sec:lambda}. On the other hand, for $\alpha\lesssim0.03$ they  can propagate as if the upstream flow were unmagnetized. The special case $\alpha=0.0$ is taken here as a representative example of the shock structure for $\alpha\lesssim0.03$.

\begin{figure*}[htb]
\begin{center}
\includegraphics[width=\textwidth]{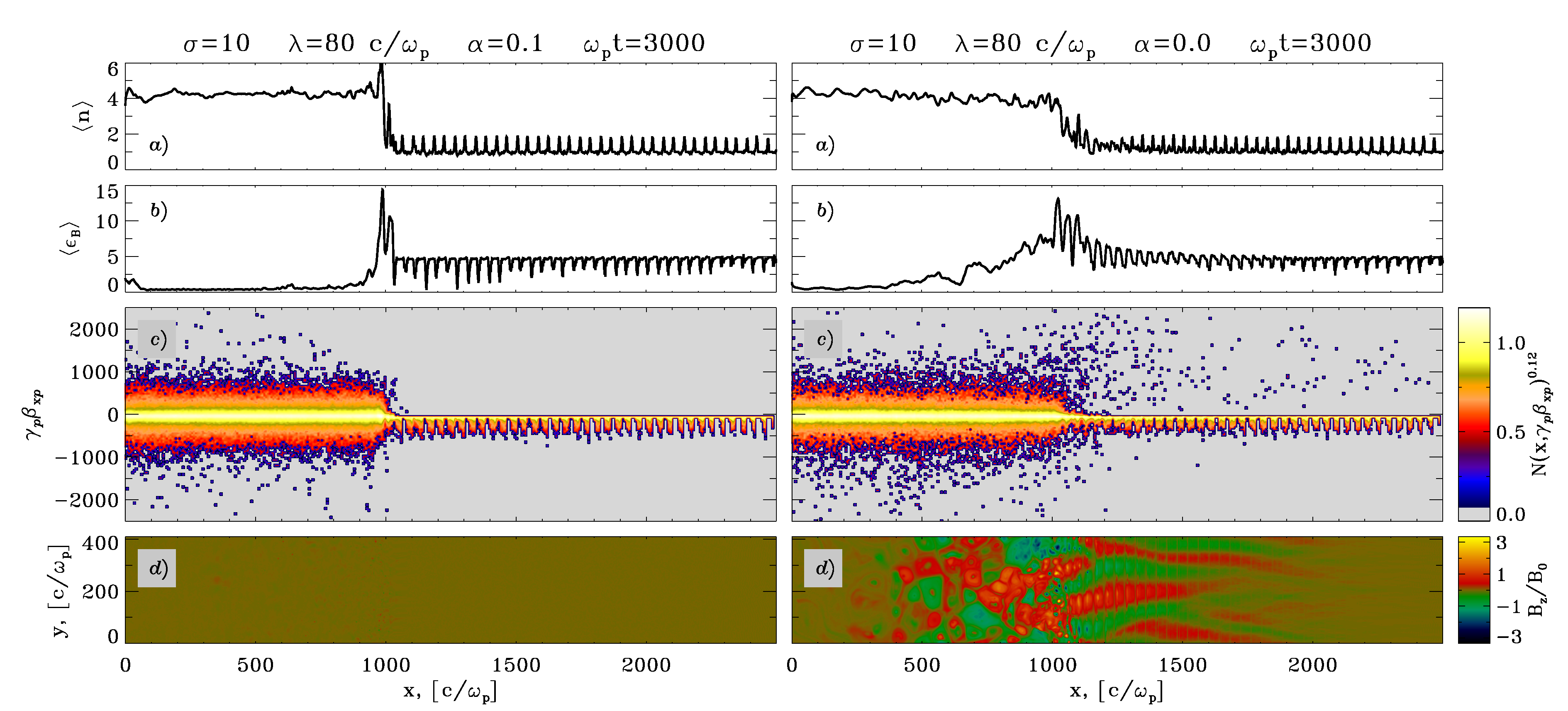}
\caption{Comparison of the shock internal structure, for a flow with $\lambda=80\comp$ and $\sigma=10$, between the case $\alpha=0.1$ (left panels) and the case $\alpha=0.0$ (right panels, corresponding to $\langle B_y\rangle_\lambda=0$). The comparison is performed at $\ompt=3000$. In both cases, the shock is located at $x\simeq1000\comp$. As a function of the longitudinal coordinate $x$, the following quantities are plotted: (a) $y$-averaged profile of the particle density $n$, in units of the pre-shock value; (b) $y$-averaged profile of the magnetic energy fraction $\epsilon_B$; (c) $x-\gb{x}$ positron phase space; (d) 2D plot on the simulation plane of the out-of-plane component $B_z$ of the magnetic field.}
\label{fig:fluiddc}
\end{center}
\end{figure*}

\fig{fluiddc} compares the structure of the flow between the cases $\alpha=0.1$ (left panels) and $\alpha=0.0$ (right panels), for the same wavelength $\lambda=80\comp$ and magnetization $\sigma=10$. The positron phase space of the case $\alpha=0.0$ (panel (c), right column) shows a diffuse population of hot particles moving back from the shock into the upstream ($x\gtrsim1050\comp$), which were absent in the case $\alpha=0.1$ (panel (c),  left column). For both $\alpha=0.0$ and $\alpha=0.1$, as the flow propagates from the fast shock (at $x\simeq1050\comp$) to the hydrodynamic shock (at $x\simeq1000\comp$), magnetic reconnection transfers energy from the field to the particles, which in the downstream region end up with an average Lorentz factor $\langle\gamma\rangle\simeq\gamma_0(\sigma+1)$ (see \S\ref{sec:lambda}). In other words, the physics of magnetic reconnection is not sensitive to the value of $\alpha$, in the regime $\alpha\lesssim0.1$. However, it is only in the case $\alpha=0.0$ (or $\alpha\lesssim0.03$, generally speaking), when the upstream stripe-averaged fields are small, that the hot particles resulting from shock-driven reconnection can stream for large distances ahead of the shock, as observed in the phase space of \fig{fluiddc}(c) (right panel). 

When a sufficient number of returning particles builds up in the pre-shock region, their counter-streaming with respect to the incoming flow can trigger the filamentation (or Weibel) instability \citep{weibel_59, medvedev_loeb_99, gruzinov_waxman_99}. In fact, the filamentary structures seen ahead of the shock in the 2D plot of $B_z$ (panel (d),  right column) are generated by the Weibel instability. A similar pattern appears in the 2D plots of electric field $E_y$ ($=-\beta_0B_z$), particle density, and magnetic energy (not shown). In contrast, no signatures of filamentation are seen for $\alpha=0.1$ (panel (d), left column), since in this case no particles can propagate far enough ahead of the shock to seed the Weibel instability.

For both $\alpha=0.0$ and $\alpha=0.1$, the hydrodynamic shock satisfies the jump conditions appropriate for an unmagnetized 3D relativistic fluid. However, the filamentation instability appreciably changes the structure of the flow for $\alpha=0.0$, especially at late times. The density profile in the transition region (panel (a)) is much smoother in the case $\alpha=0.0$ (right column) than it is for $\alpha=0.1$ (left column). More importantly, the magnetic fields generated ahead of the shock by the Weibel instability survive in the downstream region of the $\alpha=0.0$ shock, as shown by the 1D profile of magnetic energy in panel (b) (right column, at $x\lesssim1000\comp$). Instead, for $\alpha=0.1$, the only magnetic field left behind the shock comes  from compression of the pre-shock stripe-averaged field, which results in a much lower level of post-shock magnetic energy (see the left panel of \fig{fluiddc}(b)).

\begin{figure}[htb]
\begin{center}
\includegraphics[width=0.5\textwidth]{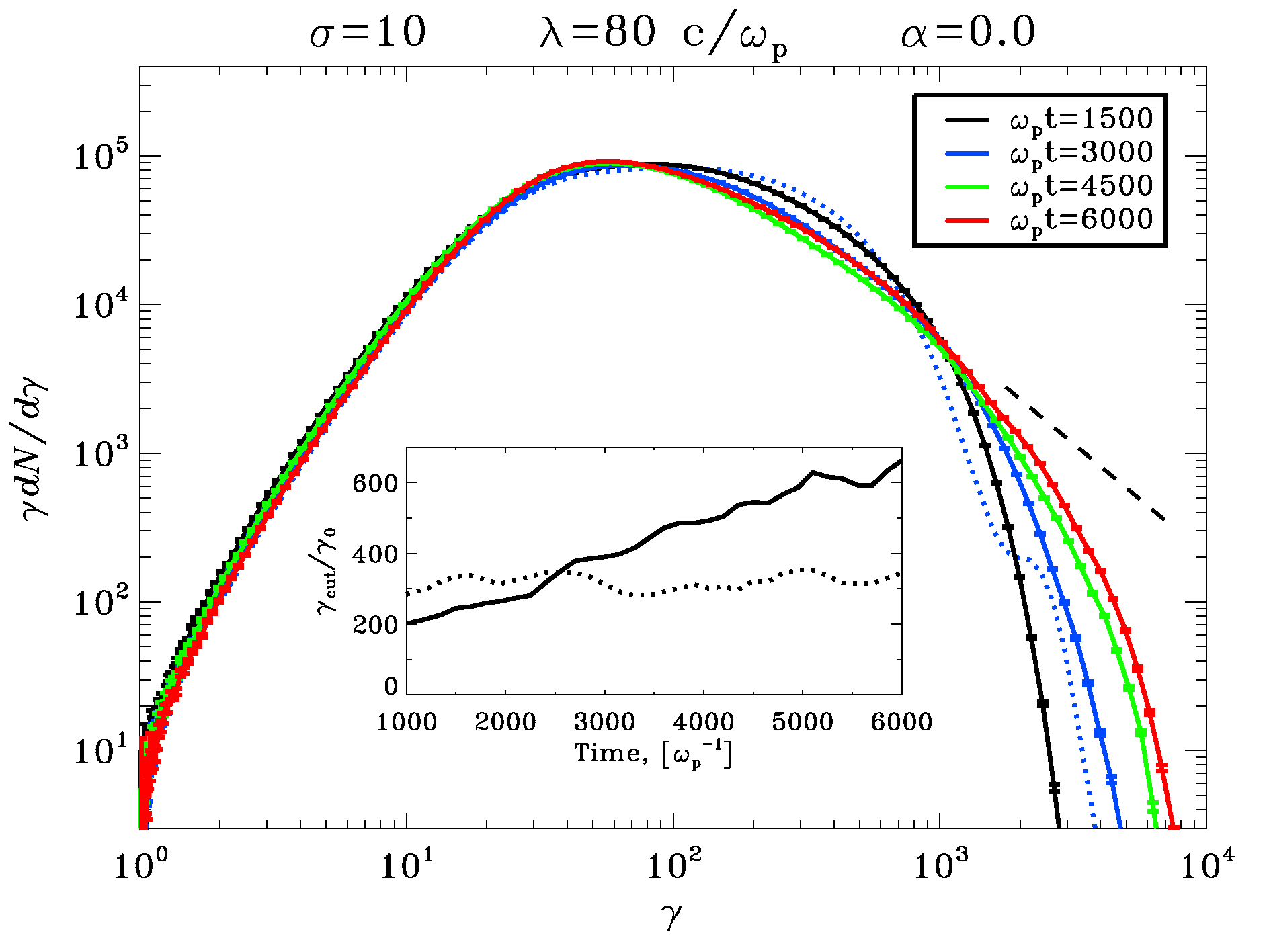}
\caption{Time evolution of the downstream spectrum for a shock propagating in a flow with $\lambda=80\comp$, $\sigma=10$, and $\alpha=0$ (i.e., zero stripe-averaged field), from $\ompt=1500$ up to $\ompt=6000$. For comparison, the downstream spectrum for a wind with the same parameters except for $\alpha=0.1$ is shown as a dotted blue line for $\ompt=3000$ (but it does not appreciably evolve in time). In the subpanel, time evolution of the maximum particle Lorentz factor, for $\alpha=0.0$ (solid line) and $\alpha=0.1$ (dotted line).}
\label{fig:spectime}
\end{center}
\end{figure}

The shape of the downstream energy spectrum is also significantly different between the cases $\alpha=0.0$ and $\alpha=0.1$. \fig{spectime} follows the time evolution of the particle distribution behind a shock with $\alpha=0.0$, compared with the case $\alpha=0.1$. As further discussed in Appendix \ref{app:time}, the downstream particle spectrum in a flow with $\alpha\gtrsim0.03$ does not appreciably change with time, so the dotted blue line in \fig{spectime}, which refers to $\ompt=3000$ for the case $\alpha=0.1$, is a good representation of the downstream energy spectrum at all times (for $\alpha=0.1$). In contrast, as shown by the solid lines, the evolution of the particle spectrum in the case $\alpha=0.0$ is much more dramatic. At early times (black line for $\ompt=1500$), the spectrum for $\alpha=0.0$ is remarkably similar to the (time-independent) particle distribution of the case $\alpha=0.1$ (dotted blue line). This confirms once more that the physics of shock-driven reconnection, which is ultimately responsible for shaping the post-shock spectrum at early times, does not appreciably depend on $\alpha$, in the regime $\alpha\lesssim0.1$. As mentioned above, the minor high-energy bump seen in the spectrum of $\alpha=0.1$ (dotted blue line), but not for $\alpha=0.0$ (black line), results from particles accelerated at the hydrodynamic shock via the SDA mechanism.

Starting from $\ompt=3000$ (solid blue line), the spectrum for $\alpha=0.0$ changes significantly. The high-energy part of the broad peak that was present at earlier times (at $\gamma\simeq300$) gets de-populated, at the expense of a prominent tail of nonthermal particles (at $\gamma\gtrsim10^3$). The tail contains approximately a few percent of particles and a few percent of the total particle energy. With time, the slope of the nonthermal tail becomes flatter (for comparison, the dashed line in \fig{spectime} has $p=2.5$), and its high-energy cutoff extends to higher Lorentz factors, as tracked by the solid line in the subpanel of \fig{spectime}. In contrast, no appreciable evolution  in the value of the maximum particle Lorentz factor is observed for the case $\alpha=0.1$ (dotted line in the subpanel). The particles in the nonthermal tail seen for $\alpha=0.0$ have been accelerated via a kind of first-order Fermi mechanism (or diffusive shock acceleration, DSA), by scattering off the turbulence generated by the Weibel instability (see \fig{fluiddc}(d), right column). Their energization proceeds in the same way as for  a strictly unmagnetized shock, which was discussed by \citet{spitkovsky_08}. By comparing the black line (for $\ompt=1500$) and the blue line (for $\ompt=3000$) at $\gamma\gtrsim10^2$, we infer that injection into the Fermi process is due primarily to the so-called ``thermal leakage,'' i.e., the hottest particles from the downstream region can escape ahead of the shock, where they take part in the Fermi acceleration cycles. Of course, once the fields produced by the Weibel instability grow stronger than the background field $B_0$ (see the scale in \fig{fluiddc}(d), right column), the Weibel-generated turbulence in the shock layer may mediate further injection of particles into the Fermi process.

Finally, we remark that, based on the condition $\lambda_{\comp}/\sqrt{\sigma}\lesssim2\, \xi_2$ discussed above, injection into the Fermi process is expected to be more favorable for short stripe wavelengths and highly magnetized winds. In fact, we observe that the efficiency of Fermi acceleration, meaning the fraction of particles participating in the Fermi cycles, is higher for smaller $\lambda$ (at fixed $\sigma$) and for larger $\sigma$ (at fixed $\lambda$). This explains why, for $\lambda=320\comp$ (so, a wavelength longer than the value $\lambda=80\comp$ used here), \fig{specdc} shows that the spectra for $\alpha=0.0$ and $\alpha=0.1$ are almost identical, in contrast with the conclusions of \fig{spectime}. For $\lambda=320\comp$, fewer particles can be injected into the Fermi process, and the structure of the flow at $\ompt=3000$ is practically the same for all values of $\alpha\lesssim0.1$. 
  

\subsection{Dependence on the Bulk Lorentz Factor}\label{sec:gamma0}
\begin{figure}[tbp]
\begin{center}
\includegraphics[width=0.5\textwidth]{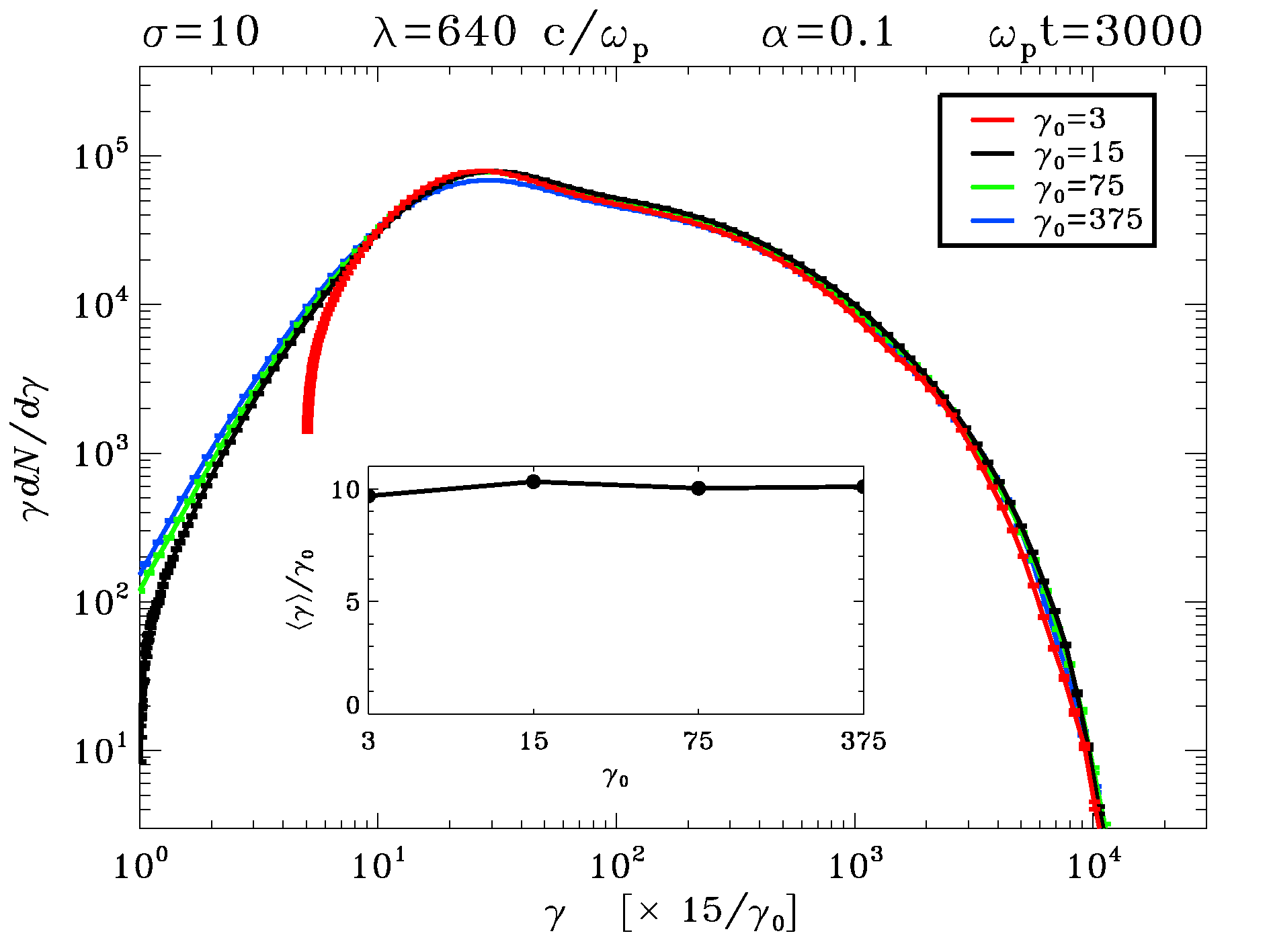}
\caption{Downstream particle spectrum at $\ompt=3000$ for different values of the pre-shock bulk Lorentz factor $\gamma_0$, in a flow with $\lambda=640\comp$, $\sigma=10$, and $\alpha=0.1$. We vary $\gamma_0$ from $\gamma_0=3$ to $\gamma_0=375$. Spectra are shifted along the $x$-axis by $15/\gamma_0$ to facilitate comparison with the reference case $\gamma_0=15$. The black line in the subpanel shows the average downstream Lorentz factor as a function of $\gamma_0$.}
\label{fig:specgam}
\end{center}
\end{figure}

\fig{specgam} shows the particle spectrum downstream from the hydrodynamic shock, for different  bulk Lorentz factors of the striped wind (from $\gamma_0=3$, in red, up to $\gamma_0=375$, in blue). The spectra are shifted along the $x$-axis by $15/\gamma_0$, to facilitate comparison with the reference case $\gamma_0=15$ (black curve) discussed in the previous sections. Once normalized in this way,  the spectra overlap almost perfectly, suggesting that the physics of shock-driven reconnection is not sensitive to the upstream bulk Lorentz factor. The value of $\gamma_0$ enters the definition of the relativistic plasma skin depth $c/\omega_{\rm p}$ ($\propto\sqrt{\gamma_0}$) and of the relativistic Larmor radius ($r_L=\comp/\sqrt{\sigma}\propto\sqrt{\gamma_0}$), but no residual dependence on $\gamma_0$ remains if length and time scales are normalized with respect to $c/\omega_{\rm p}$ and $\omega_{\rm p}^{-1}$, respectively. More specifically, the conditions that control the physics of shock-driven reconnection \citep[see \S\ref{sec:lambda}, and][]{petri_lyubarsky_07} depend only on the stripe wavelength $\lambda$ and the magnetization $\sigma$, but they do not explicitly depend on the wind Lorentz factor. This explains why our results are nearly the same (modulo an overall shift in energy scale) across a wide range of $\gamma_0$.

\section{Three-Dimensional Simulations}\label{app:3d}
The results presented so far are based on 2D experiments with magnetic field initialized in the plane of the simulations. This is the best configuration to study the tearing-mode instability, whose wavevector lies in the plane of the field, but it artificially inhibits the growth of cross-plane instabilities, which may also be important for the structure of the shock. In this section, we study the 3D physics of shock-driven reconnection, with particular emphasis on the process of particle energization. The simulation setup parallels very closely what we described in \S\ref{sec:setup}, with the magnetic field oriented initially along the $y$ direction. We adopt our fiducial values for $\sigma=10$ and $\alpha=0.1$, and we investigate two choices for the stripe wavelength, $\lambda=80\comp$ and $\lambda=160\comp$. For each value of $\lambda$, the extent of the simulation domain along $y$ and $z$ is chosen such that our results are not artificially affected by the periodicity of our boundary conditions (see Appendix \ref{app:my}). We adopt $L_y=L_z=50\comp$ for $\lambda=80\comp$, and $L_y=L_z=100\comp$ for $\lambda=160\comp$.

\begin{figure*}[tbp]
\begin{center}
\includegraphics[width=\textwidth]{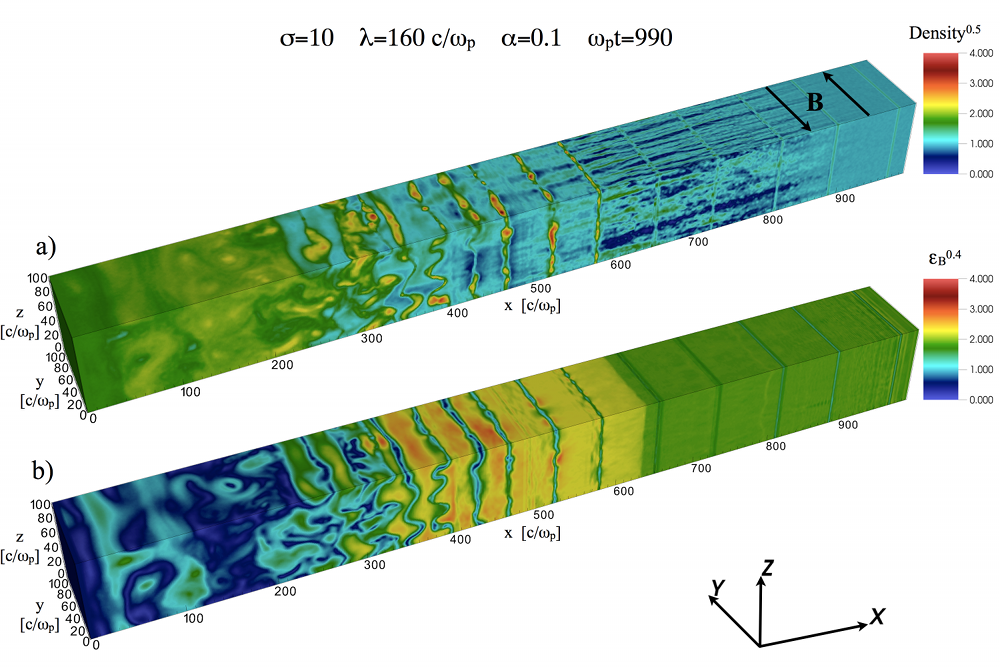}
\caption{Internal structure of the flow at $\ompt=990$, from the 3D simulation of a striped wind with $\lambda=160\comp$, $\sigma=10$, and $\alpha=0.1$. The magnetic field is initialized in the $xy$ plane, as shown by the black arrows in panel (a). The hydrodynamic shock is located at $x\simeq250\comp$, and the fast MHD shock at $x\simeq630\comp$. The following quantities are plotted: (a) 3D plot of the particle number density, in units of the upstream value, with color scale stretched to enhance contrast; (b) 3D plot of the magnetic energy fraction $\epsilon_B\equiv B^2/8 \pi \gamma_0 m n_{c0} c^2$, with color scale stretched to enhance contrast.}
\label{fig:fluid3d}
\end{center}
\end{figure*}

\fig{fluid3d} shows the structure of the shock at $\ompt=990$, for a striped wind of wavelength $\lambda=160\comp$. In agreement with the results presented in \S\ref{sec:shock} for 2D simulations, the structure of the flow in \fig{fluid3d} shows the presence of two shocks, a fast MHD shock located at $x\simeq630\comp$ and a hydrodynamic shock at $x\simeq250\comp$. At the fast shock, the flow is decelerated and compressed, as shown by the increase in magnetic energy in \fig{fluid3d}(b). The passage of the fast shock through the incoming current sheets initiates magnetic field annihilation. The dissipation of alternating fields proceeds as the flow propagates from the fast to the hydrodynamic shock, and little magnetic energy remains downstream from the hydrodynamic shock (see \fig{fluid3d}(b) at $x\lesssim250\comp$), in agreement with our 2D results. As observed in 2D, the particle density downstream from the hydrodynamic shock (\fig{fluid3d}(a)) and the shock velocity are in agreement with MHD jump conditions for a relativistic \textit{unmagnetized} fluid. In fact, most of the post-shock energy is in kinetic form, rather than electromagnetic form.\footnote{As a side note, we point out that the density filamentation seen in \fig{fluid3d}(a) ahead of the fast shock is a transient effect of the passage of the electromagnetic precursor discussed in \S\ref{sec:shock}. The precursor wave, and the resulting density filamentation, do not appreciably affect the shock structure at late times.}

As the flow propagates from the fast to the hydrodynamic shock, the structure of the current sheets is affected by the growth of two competing modes. In the $xy$ plane of the magnetic field, the tearing-mode instability breaks the current sheet into a sequence of high-density islands (see \fig{fluid3d}(a)), separated by X-points where the field lines tear and reconnect. In the $xz$ plane orthogonal to the field, the folding of the current sheet seen in \fig{fluid3d} is governed by the drift-kink instability, driven by the current of fast-drifting plasmas in a thin sheet \citep{daughton_98, zenitani_05, zenitani_07}. The 2D geometry employed in the previous sections was chosen to select the tearing mode, and suppress the drift-kink mode. In the regime of undriven reconnection, the drift-kink instability is generally thought to grow faster than the tearing mode \citep{zenitani_05b,zenitani_07}, but for very thin sheets the opposite hierarchy is observed, as shown via 3D PIC simulations by \citet{liu_11}. In our experiments of shock-driven reconnection, the current sheet just behind the fast shock ($x\simeq580\comp$) shows the pattern of magnetic islands typical of the tearing mode, whereas the drift-kink mode appears only farther downstream ($x\lesssim500\comp$). 

The nature of the instability that dominates the process of field annihilation leaves an imprint on the resulting particle distribution. \citet{zenitani_07} have shown with PIC simulations  that  field dissipation due to the drift-kink instability does not result in nonthermal particle acceleration, the plasma just being heated. This is because the field lines remain straight, so that all particles gain energy at the same rate. Nonthermal particles are produced only by the tearing mode \citep[e.g.,][]{zenitani_01}. In our 2D simulations with in-plane fields, i.e., the geometry required to capture the tearing mode, we have shown that particles are accelerated to nonthermal energies by the reconnection electric field. They initially move along the $z$ direction parallel to the reconnection field, and then they drift along $y$ from a given X-point into the closest magnetic island. In 3D, one could argue that the folding of the current sheet introduced by the drift-kink instability in the $xz$ plane (not resolved in 2D) may deflect the accelerating particles out of the current sheet, thus suppressing their energization. 

\begin{figure}[tbp]
\begin{center}
\includegraphics[width=0.5\textwidth]{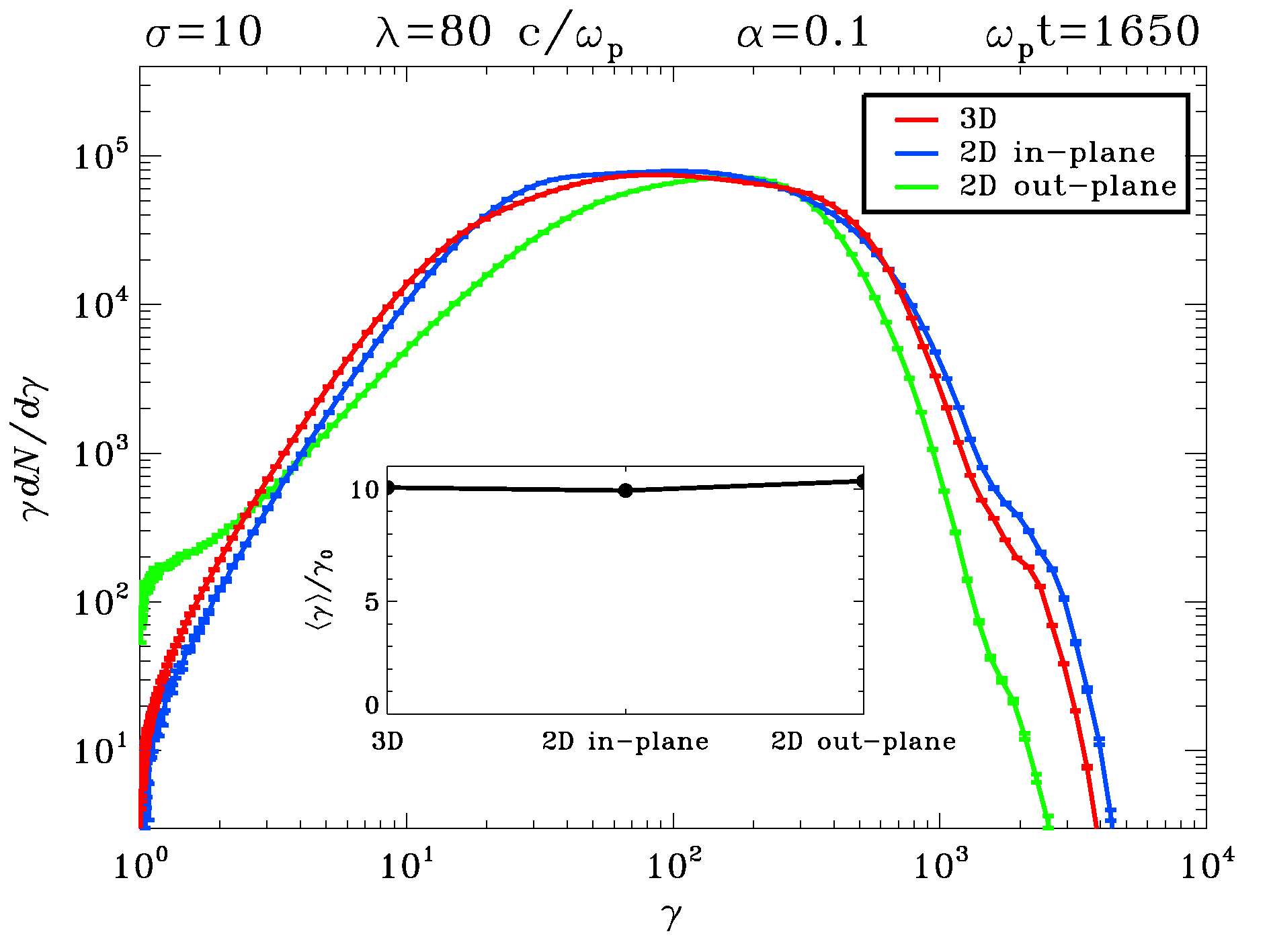}
\caption{In a striped wind with $\lambda=80\comp$, $\sigma=10$, and $\alpha=0.1$, comparison of the downstream particle spectrum for different magnetic field configurations: 3D simulation (red) and 2D simulations with either in-plane (blue) or out-of-plane (green) magnetic fields. The black line in the subpanel shows the average downstream Lorentz factor  $\langle\gamma\rangle$ for the different cases.}
\label{fig:spec3d}
\end{center}
\end{figure}

In \fig{spec3d}, we address this important issue, by comparing for $\lambda=80\comp$ the post-shock particle spectrum of a 3D simulation (red line) to the results of 2D experiments with in-plane or out-of-plane magnetic fields (blue and green lines, respectively). The excellent agreement between 3D and 2D in-plane results (red and blue line, respectively) suggests that the physics of particle acceleration by shock-driven reconnection is captured extremely well by our 2D simulations with in-plane fields.\footnote{The agreement between red and blue lines in the high-energy bump at $\gamma\gtrsim2000$ also suggests that the physics of SDA is correctly described by our 2D simulations with in-plane fields.} In turn, this implies that, at least for $\lambda=80\comp$, the maximum energy of accelerated particles is constrained by the distance between X-points and islands (in the $xy$ plane), rather than by the current sheet folding in the $xz$ plane. Just upstream of the hydrodynamic shock, the typical spacing between X-points and islands scales linearly with the stripe wavelength. A similar trend may be expected in the characteristic scale of the drift-kink mode, for the current sheet just ahead of the hydrodynamic shock (see \fig{fluid3d}(b) at $x\simeq350\comp$). If this is the case, our conclusions will also hold for longer stripe wavelengths, as we have directly verified in the case  $\lambda=160\comp$. A detailed 3D study of particle acceleration by shock-driven reconnection will be presented elsewhere.

Finally, we point out that the tension force of a guide field (along $z$, in our geometry) can easily stabilize the drift-kink mode, with little or no effect on the tearing mode  \citep{zenitani_08}. In the presence of a guide field, the 3D physics of shock-driven reconnection should be described very accurately by 2D simulations with alternating fields lying in the simulation plane. A guide field component, though, is not usually expected in the context of pulsar winds \citep[but see][]{petri_kirk_05}.


\section{Summary and Discussion}\label{sec:disc}
We have explored by means of 2D and 3D PIC simulations the internal structure and acceleration properties of relativistic shocks that propagate in an electron-positron striped wind, i.e., a flow consisting of stripes of alternating field polarity separated by current sheets of hot plasma. This work extends the study of \citet{sironi_spitkovsky_09}, that investigated relativistic shocks in a pair plasma carrying uniform magnetic fields. 

We find that a fast MHD shock propagates into the pre-shock striped flow, compressing the incoming current sheets and initiating the process of \textit{driven magnetic reconnection}, via the tearing-mode instability. Reconnection islands seeded by the passage of the fast shock grow and coalesce, while magnetic energy is dissipated at X-points located in between each pair of islands. When reconnection islands grow so big to occupy the entire region between neighboring current sheets, the striped structure of the flow is erased, and a hydrodynamic shock forms. Downstream from the shock, the average particle energy is larger than in the pre-shock flow by a factor of $\simeq\sigma$, where $\sigma$ is the wind magnetization. In other words, the energy stored in the alternating fields has been entirely transferred to the particles via shock-driven magnetic reconnection, and the downstream fluid behaves like an unmagnetized plasma. The only field component surviving in the downstream region comes from shock-compression of the stripe-averaged pre-shock field (i.e., the non-alternating component), if non-zero.  In our 2D and 3D simulations, we find that complete annihilation of the alternating fields occurs irrespective of the stripe wavelength $\lambda$ or the magnetization $\sigma$, in contrast with the results of the 1D analysis by \citet{petri_lyubarsky_07}. In Appendix \ref{app:my} we show that multi-dimensional studies are of paramount importance to correctly capture the physics of shock-driven reconnection. 

The main agent of particle energization is the reconnection electric field, as particles drift from a given X-point into the closest island. Whether all particles have comparable energy gains, or only a few particles are accelerated to high energies, and the majority stay cold, depends sensitively on the properties of the wind. We have explored the dependence of the downstream particle spectrum on the stripe wavelength $\lambda$, the wind magnetization $\sigma$ (or equivalently, the field strength $B_0$), and the stripe-averaged magnetic field $\langle B_y\rangle_\lambda$ (or equivalently, the parameter $\alpha$, with $\alpha=0$ for $\langle B_y\rangle_\lambda=0$, and $|\alpha|\rightarrow1$ in the limit $|\langle B_y\rangle_\lambda|\rightarrow B_0$). For fixed $\alpha=0.1$, we find that the shape of the post-shock spectrum depends primarily on the combination $\lambda/(r_L\sigma)$, where $r_L$ is the relativistic Larmor radius in the striped wind. This is just the stripe wavelength measured in units of the post-shock plasma skin depth. For small values of $\lambda/(r_L\sigma)$ ($\lesssim$ a few tens), the spectrum resembles a Maxwellian distribution with mean energy $\langle\gamma\rangle\simeq\gamma_0\sigma$, where $\gamma_0$ is the bulk Lorentz factor of the pre-shock flow. In the limit of very large values of $\lambda/(r_L\sigma)$ ($\gtrsim$ a few hundreds), the spectrum approaches a broad power-law tail, with a flat spectral slope $p\simeq1.5$. In this regime, the tail extends from  $\gamma_{\rm min}\simeq\gamma_0$ up to $\gamma_{\rm max}\simeq \gamma_0 \sigma^{1/(2-p)}$. In terms of the flow structure, the former case (i.e., a Maxwellian-like spectrum) is realized when all particles pass in the vicinity of an X-point, thus gaining energy from field dissipation. In contrast, the latter case occurs when most of the particles remain far from X-points, thus staying cold, and the energy of annihilating fields is transferred to only a few particles, which end up dominating the energy content of the flow.

As $|\alpha|$ increases from $0.1$ up to $1$, the fraction of upstream Poynting flux in the form of alternating fields becomes smaller. As the amount of field energy available for dissipation decreases, the average particle Lorentz factor downstream from the shock recedes from $\langle\gamma\rangle\simeq\gamma_0\sigma$, as expected for complete field annihilation, down to $\langle\gamma\rangle\simeq\gamma_0$, the result of  an unstriped wind. 
In the opposite limit $|\alpha|\lesssim0.1$, i.e., in the case of nearly symmetric stripes, the particles accelerated by the reconnection electric field to the highest energies can escape ahead of the shock, unaffected by the stripe-averaged field that would tend to advect them back downstream. Their counter-streaming with respect to the incoming flow can seed the filamentation (or Weibel) instability, as it happens for shocks propagating in unmagnetized plasmas \citep{spitkovsky_05,spitkovsky_08}. In turn, the turbulence generated by the Weibel instability can mediate particle acceleration to even higher energies, via a Fermi-like diffusive process. The nonthermal tail produced by the Fermi mechanism is steeper than that given by magnetic reconnection, with characteristic slopes $p\gtrsim2.5$. Typically, a few percent of particles are injected into the Fermi process, and the injection efficiency scales in inverse proportion to the ratio $\lambda/(r_L\sigma)$. With time, the nonthermal tail of Fermi-accelerated particles extends to higher energies. For $|\alpha|\gtrsim0.1$, the propagation of particles far ahead of the shock is inhibited by the stripe-averaged field, and Fermi acceleration is suppressed. In this case, a few percent of particles can get accelerated by the stripe-averaged electric field $\langle E_z\rangle_\lambda$ while gyrating around the shock front, a process known as shock-drift acceleration \citep[i.e.,][]{begelman_kirk_90}. The tail resulting from shock-drift acceleration has a limited energy extent, and it does not evolve in time.

We have confirmed our findings with an extensive set of convergence tests. Most importantly, as we have shown in \S\ref{app:3d}, 2D simulations with in-plane magnetic fields -- the setup that we have primarily adopted in this work -- can reproduce very accurately the shock physics and particle spectrum observed in 3D simulations. In other words, the current sheet folding caused by the drift-kink instability in the plane orthogonal to the field \citep{zenitani_07} does not represent a serious constraint for the maximum particle energy achieved in the course of the reconnection process. Rather, the maximum energy of accelerated particles is limited by the time to drift from a given X-point into the closest island.


These findings can place important constraints on current models of Pulsar Wind Nebulae (PWNe), bubbles of synchrotron-emitting plasma powered by the relativistic wind of young pulsars. If the magnetic and rotation axes of the pulsar are not aligned, the wind consists of stripes of opposite magnetic field polarity, alternating with the pulsar period. The wind wavelength will be $\lambda=2\pi R_{LC}$, where $R_{LC}=c/\Omega$ is the so-called light-cylinder radius, and $\Omega$ is the angular frequency of the pulsar. The stripe-averaged field vanishes along the equatorial plane (i.e., $\alpha=0$ there), and the striped structure persists up to a latitude equal to the inclination angle between the pulsar's magnetic and rotational axes. At higher latitudes, the wind carries a uniform (i.e., non-alternating) field, so the physics discussed in the present work does not apply \citep[see, instead,][]{gallant_92, hoshino_92, amato_arons_06, sironi_spitkovsky_09, sironi_spitkovsky_11a}.

Our results suggest that efficient dissipation of the alternating fields should occur at the termination shock of pulsar winds, irrespective of the properties of the flow. It follows that current observational constraints on the upstream magnetization parameter \citep{rees_gunn_74, kennel_coroniti_84, delzanna_04} should be attributed not to the total Poynting flux, but to the Poynting flux associated with the stripe-averaged magnetic field, the only component surviving downstream from the shock. The upstream flow may be Poynting-dominated, provided that most of the Poynting flux is in the form of an alternating field, which is annihilated at the shock. Thus, our results provide a solution to the so-called ``sigma problem,'' i.e., that pulsar winds should be magnetically-dominated at the light-cylinder radius, but kinetically-dominated downstream from the termination shock. The idea of shock-driven reconnection as a solution to the sigma problem was originally proposed by \citet{lyubarsky_03}. Our work demonstrates that complete annihilation of the alternating fields occurs regardless of the properties of the wind, in the regime of relativistic magnetically-dominated flows.

In addition, the particle energy spectra extracted from our simulations can be used directly to interpret the radiative signature of PWNe. The radio spectrum of the Crab Nebula, the prototype of the class of PWNe, requires a population of nonthermal particles with a flat spectral slope ($p\simeq1.5$), extending at least across three decades in energy, from $10^2\unit{MeV}$ up to $10^5\unit{MeV}$. The particle spectrum should be steeper at higher energies, with $p\simeq2.5$, to explain the optical and X-ray flux.

Our results suggest that, for a slope $p\simeq1.5$, the width of the electron power-law tail resulting from reconnection will be $\simeq\sigma^2$. To produce a particle distribution extending over three decades in energy, the wind magnetization at the termination shock needs to be $\sigma\gtrsim30$. Most importantly, broad particle spectra are produced by shock-driven reconnection only if $\lambda/(r_L \sigma)\gtrsim$ a few tens. For pulsar winds, the particle number density scales with distance from the pulsar as $n=n_{LC}(R_{LC}/R)^2$, where  the density at the light-cylinder radius is usually written as $n_{LC}=\kappa\,n_{GJ}$. Here, $n_{GJ}=\Omega B_{LC}/2\pi ec$ is the Goldreich-Julian density \citep{goldreich_julian_69}, and $\kappa$ is the so-called multiplicity. Conservation of energy along the flow streamlines implies that at the termination shock $\gamma_0 (1+\sigma) \kappa=\omega_{LC}/2\,\Omega$, where $\omega_{LC}=eB_{LC}/mc$. It follows that at the termination shock ($R=R_{TS}$) 
\be\label{eq:wind}
\frac{\lambda}{r_L \sigma}\simeq4\pi\kappa \frac{R_{LC}}{R_{TS}}~~.
\ee
The constraint $\lambda/(r_L \sigma)\gtrsim$ a few tens required to produce broad particle spectra is extremely challenging for current models of PWNe and pulsar magnetospheres. The radius of the Crab termination shock in the equatorial plane is inferred from X-ray observations to be $R_{TS}\simeq0.1\unit{pc}\simeq5\times10^8R_{LC}$ \citep{hester_02}. Most available models estimate $\kappa\simeq10^4-10^6$ \citep{bucciantini_11}, depending on whether or not radio-emitting electrons are included in the analysis. Based on our findings,  the resulting value of $\lambda/(r _L \sigma)\lesssim0.01$ would yield a Maxwellian-like spectrum, at odds with the wide flat spectrum required by observations.

If radio-emitting electrons are produced in the equatorial plane by shock-driven reconnection, a revision of the existing estimates of $\kappa$ is required. In this respect, we point out that the values of $\kappa$ quoted above are averages over latitude, and one cannot exclude that particle injection into the pulsar wind is highly anisotropic, with multiplicity as large as $10^8$ along the equatorial plane. Of course, any given choice of $\kappa$ will constrain $\gamma_0$ and $\sigma$ via  $\gamma_0 (1+\sigma) \kappa=\omega_{LC}/2\,\Omega$, a relation that holds at each latitude. From the equatorial plane, radio-emitting electrons will quickly fill the whole nebula, transported by the strong fluid motions observed in MHD models of PWNe downstream from the termination shock \citep[e.g.,][]{komissarov_04, delzanna_04, camus_09}. This would provide a natural explanation for the lack of gradients in the radio spectral slope of the Crab \citep{bietenholz_kronberg_92}. Alternatively, the radio part of the spectrum may be produced at higher latitudes, where the termination shock is closer to the pulsar, and the ratio in \eq{wind} becomes larger. However, in this case one should also account for the fact that, at high latitudes, the fraction of upstream Poynting flux available for dissipation is lower (since $|\alpha|$ is larger), which results in a narrower spectrum, everything else being fixed.\footnote{Also, the termination shock at high latitudes is oblique with respect to the wind velocity, a configuration not studied here.}

On the other hand, the small value of $\lambda/(r_L \sigma)$ expected in the equatorial plane on the basis of current models of the Crab Nebula is the most favorable setup for particle acceleration via the Fermi diffusive process, with efficiency of a few percent by number. In this respect, one could interpret the optical and X-ray signature of the Crab, which requires a particle spectrum with $p\simeq2.5$, as synchrotron emission from such Fermi-accelerated particles. Based on our results, optical and X-ray emitting electrons should be produced close to the equatorial plane of the wind, where $|\alpha|\lesssim0.1$. If the same equatorial wedge is responsible for accelerating the radio-emitting electrons, our findings predict that they should outnumber the optical and X-ray emitting particles by a factor of few hundred. This is in agreement with current models of PWNe spectra, which require $\kappa\simeq10^6$ if the radio band is included in the analysis \citep{bucciantini_11}, a value which is two orders of magnitude larger than if only higher frequency bands are considered \citep[$\kappa\simeq10^4$,][]{kennel_coroniti_84b}.

Although our work focuses primarily on the physics of PWNe, shocks propagating into striped flows may be present also in blazar jets and gamma-ray bursts \citep[e.g.,][]{thompson_06}, where they may provide an appealing explanation, based on strong physical grounds, for any flat particle distribution inferred from observations. The broad-band emission of hotspots in radio galaxies, usually interpreted as the termination shocks of relativistic jets, indicates that the spectra of accelerated electrons need to be flat ($1<p<2$) below GeV energies \citep{stawarz_07}. In the case of luminous blazar sources, very flat electron spectra below GeV energies are inferred directly from X-ray observations \citep{celotti_08, sikora_09}. The steeper particle spectrum required at $\gtrsim$GeV energies, with slope $p\simeq2.5$, could instead result from Fermi acceleration across the shock, if the stripe-averaged field satisfies $|\alpha|\lesssim0.1$. 

\acknowledgements
We thank J.~Arons,  R.~Blandford, N.~Bucciantini, A.~Loeb, J.~McKinney, R.~Narayan, and E.~Quataert for insightful comments. L.S. gratefully thanks  D.~Giannios for many inspiring discussions. This work was supported by NSF
grant AST-0807381 and NASA grant NNX10AI19G. The simulations presented in this
article were performed on computational resources supported
by the PICSciE-OIT High Performance Computing
Center and Visualization Laboratory, and at  National
Energy Research Scientific Computing Center, which is
supported by the Office of Science of the US Department
of Energy under contract No. DE-AC02-05CH11231.

\appendix

\section{(A) Dependence on the Transverse Size of the Simulation Box}\label{app:my}
\begin{figure}[htbp]
\begin{center}
\includegraphics[width=0.5\textwidth]{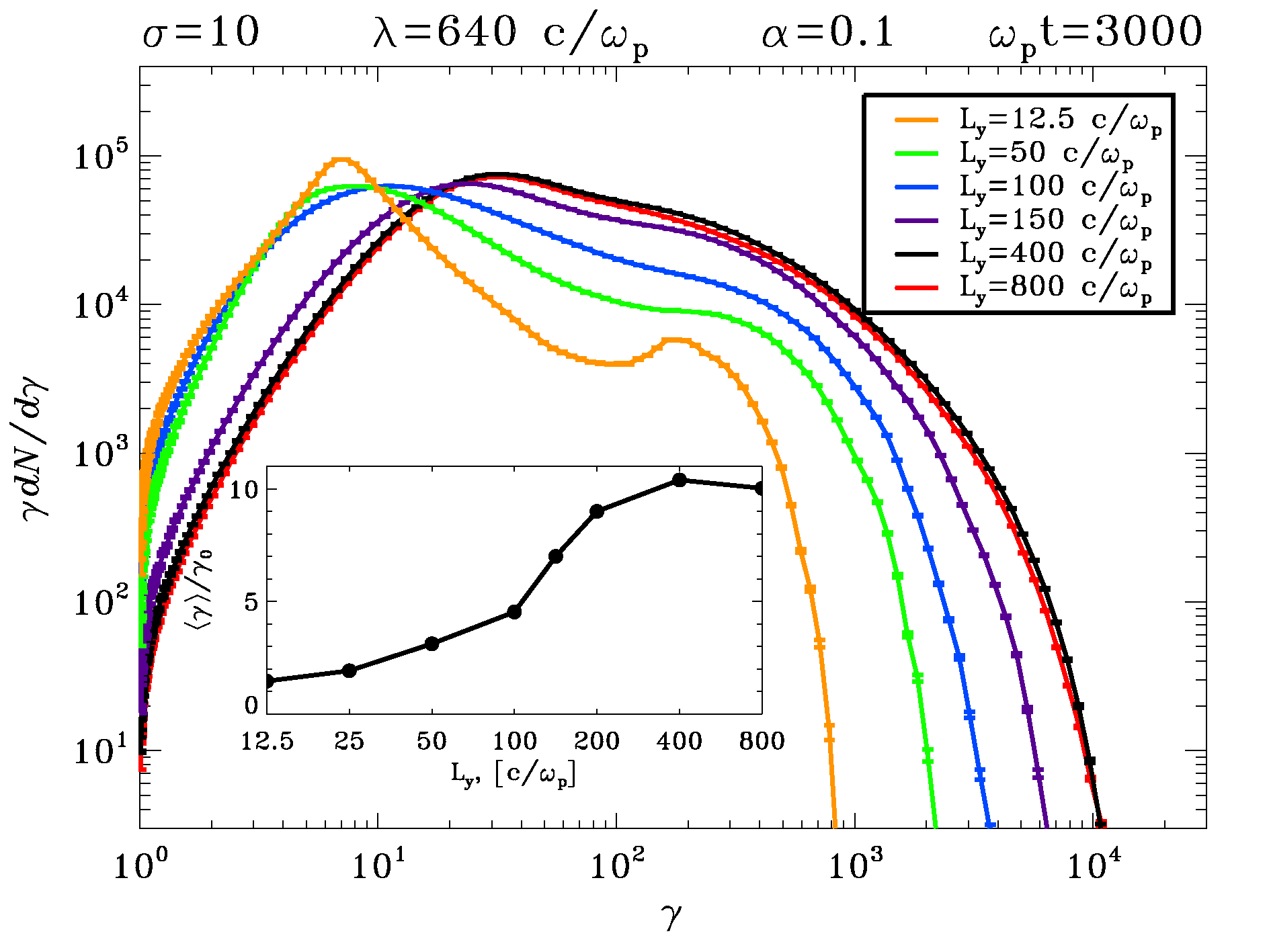}
\caption{Downstream particle spectrum at $\ompt=3000$ for different values of the transverse size $L_y$ of our simulation box, in a flow with $\lambda=640\comp$, $\sigma=10$, and $\alpha=0.1$. We vary $L_y$ from $L_y=800\comp$ down to $L_y=12.5\comp$, which approaches the 1D setup of \citet{petri_lyubarsky_07}. The black line in the subpanel shows the average downstream Lorentz factor  $\langle\gamma\rangle$ as a function of the box width $L_y$.}
\label{fig:specmy}
\end{center}
\end{figure}

In \fig{specmy} we investigate the dependence of our findings on the transverse size $L_y$ of the computational domain, for $\lambda=640\comp$, $\sigma=10$, and $\alpha=0.1$. The agreement between the black curve ($L_y=400\comp$) and the red line ($L_y=800\comp$) suggests that our results become insensitive to the transverse size of the box (i.e., they converge with respect to $L_y$), for boxes larger than $400\comp$. Generally speaking, we find that a good criterion for convergence is $L_y\gtrsim\lambda/2$. When this condition is fulfilled, the growth of magnetic islands behind the fast shock is not artificially inhibited by the periodicity of our boundaries in the $y$ direction. Magnetic reconnection then  proceeds up to the point when islands from two neighboring current sheets will merge, which happens when their size is $\simeq\lambda/2$ (for $\alpha\lesssim0.1$).

On the other hand, for $L_{y}\lesssim 300\comp$ the growth and coalescence of reconnection islands within a given current sheet artificially stops when only one island is left in the sheet, and its size approaches $L_y$. At this point, a second shock forms, located behind the fast shock. For $L_y\gtrsim400\comp$, this would correspond to the hydrodynamic shock discussed in \S\ref{sec:shock}, where the striped structure of the flow is erased, and field energy is entirely transferred to the particles. Instead, in the case $L_{y}\lesssim 300\comp$, complete dissipation of the alternating fields is prohibited by the fact that the growth of islands is artificially inhibited by the transverse extent of the box. Here, the post-shock flow retains a striped structure, with regions of hot plasma separated by highly magnetized walls of cold particles. With decreasing $L_y$, less magnetic energy is transferred to the particles behind the fast shock, and the post-shock flow gets more dominated by magnetic field energy, with respect to particle kinetic energy (see the decrease in average particle energy with decreasing $L_y$, in the subpanel of \fig{specmy}). This also explains why in \fig{specmy} the normalization of the low-energy peak, populated by the cold particles in the high-field regions, grows with decreasing $L_y$, at the expense of the high-energy component of hot particles that gained energy from field dissipation. 

As discussed in the main body of the paper (see  \S\ref{sec:lambda}), the maximum size of magnetic islands (which here is set by the smallest between $\lambda/2$ and $L_y$) correlates with the highest energy at which particles can be accelerated by the reconnection electric field. The trend in the upper spectral cutoff of \fig{specmy} can then be explained by the decrease with $L_y$ in the maximum size of reconnection islands. In this respect, the spectra in \fig{specmy}, interpreted as a monotonic trend in the size of reconnection islands, can be directly compared to the spatial sequence of spectra in \fig{fluidsh}(f), which follow the growth of magnetic islands as the flow propagates from the fast to the hydrodynamic shock. A similar trend as in \fig{specmy} is also observed in the large-$\alpha$ limit of \fig{specdc} (yet for a different wavelength, $\lambda=320\comp$ as opposed to $\lambda=640\comp$). However, in that case the maximum size of magnetic islands was set by a physical constraint, namely the smallest between $\lambda_+$ (where $B_y=+B_0$) and $\lambda_-$ (where $B_y=-B_0$), whereas here it is the transverse size of the box which artificially limits  the growth of magnetic islands for $L_y\lesssim300\comp$.

As the downstream region gets more magnetized with decreasing $L_y$, the main shock, which would be moving at $\beta_{\rm sh}\simeq1/3$ for $L_y\gtrsim400\comp$, propagates at a faster velocity, eventually catching up with the fast MHD shock, in the limit $L_y\ll\lambda$. In this regime, we recover the 1D results by \citet{petri_lyubarsky_07}. The wind parameters employed in this section, $\lambda=640\comp$ and $\sigma=10$, satisfy the condition $\lambda_{\comp}/\sigma\gtrsim8\,\xi_1/5$ (where $\xi_1=6-10$). In this case, the 1D model by  \citet{petri_lyubarsky_07} would predict negligible field dissipation, in agreement with our results for a very narrow box (yellow line for $L_y=12.5\comp$; see also the subpanel, for small values of $L_y$). However, as clarified by  \fig{specmy}, this is just an artificial consequence of the reduced dimensionality of their model, which cannot correctly capture the development of the tearing-mode instability, and the resulting growth and coalescence of magnetic islands. In a more realistic 2D scenario, complete field dissipation is achieved by the time the flow enters the hydrodynamic shock. This stresses that multi-dimensional simulations are essential for our understanding of shock-driven reconnection.

\section{(B) Time Evolution of the Shock}\label{app:time}
\begin{figure*}[htbp]
\begin{center}
\includegraphics[width=\textwidth]{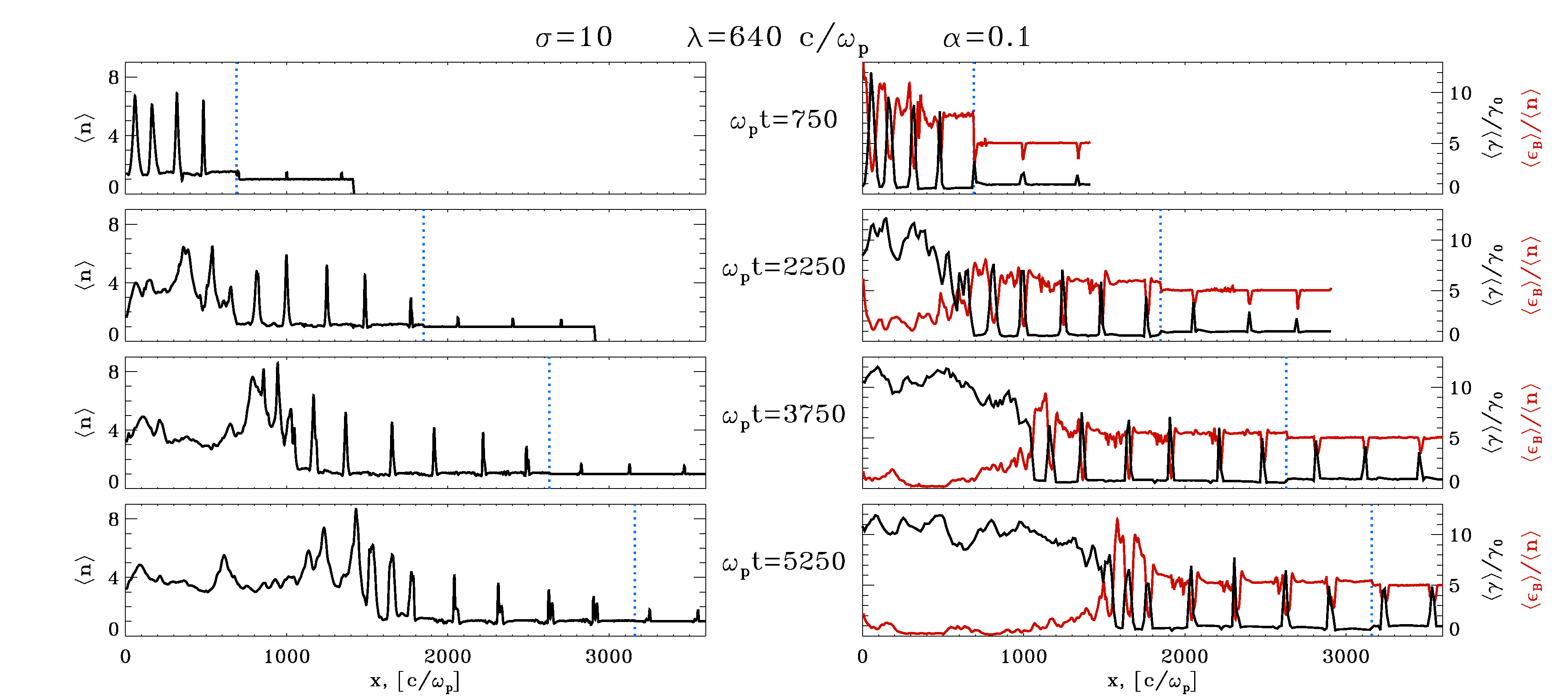}
\caption{Time evolution of the internal structure of a shock propagating in a striped flow with $\lambda=640\comp$, $\sigma=10$, and $\alpha=0.1$. We follow the evolution of the shock from $\ompt=750$ up to $\ompt=5250$ (from top to bottom). Left panels show the $y$-averaged particle number density, in units of the upstream value. Right panels show the mean kinetic energy per particle (black lines) and the mean magnetic energy per particle (red lines). In all panels, the location of the fast MHD shock is indicated as a vertical dotted blue line.}
\label{fig:fluidtime}
\end{center}
\end{figure*}

In \fig{fluidtime} we follow the time evolution of the shock, from the earliest stages until it reaches a steady state. We choose our fiducial values for $\lambda=640\comp$, $\sigma=10$, and $\langle B_y\rangle_\lambda\simeq0.05$ (corresponding to $\alpha=0.1$), but our main conclusions hold basically for the whole parameter space we have investigated. At very early times (uppermost row, for $\ompt=750$), only the fast MHD shock is present (its location at different times is shown as a vertical dotted blue line in \fig{fluidtime}). The incoming flow decelerates at the fast shock, forming a ring in momentum space lying in the $xz$ plane orthogonal to the magnetic field. Since the spread in $y$-momentum is much smaller than in the other two directions, the fluid behind the fast shock can be treated as a 2D relativistic plasma. In fact, the increase in density, magnetic energy, and transverse magnetic field across the fast shock, as well as the drop in average  kinetic energy per particle, are in good agreement with the MHD jump conditions for a 2D relativistic fluid. 

Behind the fast shock, the energy content of the flow at $\ompt=750$ is dominated by magnetic fields. However, even at such early times, we see that current sheets are getting wider (left panel) and populated with hotter particles (see the mean kinetic energy per particle, black line in the right panel), as we proceed farther downstream from the fast shock. As described in \S\ref{sec:struct}, a larger and larger fraction of the cold wind is being channeled into reconnection islands, seeded by the passage of the fast shock through the upstream current sheets. At X-points located in between each pair of neighboring islands, magnetic energy is being transferred to the particles, which explains why the peaks in mean \tit{kinetic} energy per particle are getting higher with distance behind the fast shock (black line in the right panel), at the expense of the mean \tit{magnetic} energy per particle (red line). Yet, the striped structure, with  magnetized regions of cold plasma separated by hot current sheets, still persists downstream from the fast shock.

The shock structure changes significantly at $\ompt=2250$  (second row from the top). Behind the fast shock, which is now much weaker (see the modest density jump at $x\simeq1850\comp$), reconnection islands have enough time to grow and fill the entire region in between neighboring current sheets. At $x\simeq700\comp$, the striped structure is erased, and a hydrodynamic shock forms (see the density profile in the left panel). The density plateau behind the hydrodynamic shock  is in agreement with the expected jump conditions for a relativistic unmagnetized 3D plasma, namely $n_{\rm d}/n_{\rm u}\simeq4$. In fact, the energy balance presented in the right panel shows that, by the time the flow crosses the hydrodynamic shock, most of the field energy (red line) has been converted to particle kinetic energy (black line).  

A similar picture holds for later times ($\ompt=3750$ in the third row from the top, $\ompt=5250$  in the lowermost row). The density behind the hydrodynamic shock stabilizes at $n_{\rm d}\simeq4\,n_{\rm u}$, and the transfer of energy from magnetic to kinetic form becomes more and more complete. The change of adiabatic index due to the transition from a magnetically-dominated fluid to a kinetically-dominated plasma results in a lower pressure to drive the fast shock, that decelerates (compare the locations of the vertical dotted blue lines at different times). Between $\ompt=3750$ and $\ompt=5250$, the overall structure of the transition region reaches a steady state. The fast shock propagates ahead of the hydrodynamic shock, roughly  at a fixed distance (i.e., the speed of the fast shock approaches the velocity $\beta_{\rm sh}\simeq1/3$ of the hydrodynamic shock). The separation between the two shocks obeys the requirement that reconnection islands seeded by the fast shock should fill the entire region in between current sheets, by the time the flow arrives at the hydrodynamic shock. When the flow structure has reached the steady state, we find that the particle spectrum downstream from the hydrodynamic shock does not significantly change with time.

\bibliography{stripe}
\end{document}